\documentclass[sigconf, 10pt, nonacm]{acmart}
\makeatletter
\def\@ACM@checkaffil{}
\makeatother

\AtBeginDocument{%
  }

\author{Anton Juerss}
\affiliation{
  \institution{Weizenbaum Institute \& TU Berlin}
}

\author{Stefan Schmid}
\affiliation{
  \institution{TU Berlin \& Weizenbaum Institute}
}

\usepackage{algorithm}
\usepackage{algorithmic}

\usepackage{float}
\usepackage{enumitem}

\usepackage{xspace}

\newcommand{\name}{\textsc{Bridge}\xspace}
\newcommand{\gbruck}{\textsc{G-Bruck}\xspace}
\newcommand{\bruck}{\textsc{Bruck}\xspace}
\newcommand{\sbruck}{\textsc{S-Bruck}\xspace}
\newcommand{\rehd}{\textsc{R-HD}\xspace}
\newcommand{\halving}{\textsc{Halving-Doubling}\xspace}
\newcommand{\hd}{\textsc{HD}\xspace}
\newcommand{\ring}{\textsc{Ring}\xspace}

\definecolor{cbBlue}{RGB}{0,114,178}

\usepackage{subcaption}
\begin{document}

\title[\textsc{Bridge}: Collective Communication Schedules with Reusable Subrings]{\name: Optimizing Collective Communication Schedules in Reconfigurable Networks with Reusable Subrings}

\begin{abstract}
Optical circuit-switched networks have emerged as an appealing alternative to electrical fabrics as they can reconfigure the network topology at runtime, reducing communication cost and improving bandwidth utilization. Yet exploiting optical reconfigurable networks for collective communication comes with a fundamental trade-off: each reconfiguration incurs non-negligible delay and the benefit of a new topology depends on future traffic. The central question is therefore when reconfiguration is worth its cost. While prior work has demonstrated the benefits of reconfiguration, the opportunity to \textit{reuse} optical links to optimize multiple steps remains largely unexplored.

In this paper, we present \name, a reconfiguration strategy for frequently used collective communication primitives, namely All-to-All, AllReduce, Reduce-Scatter, and AllGather. \name exploits the structure of Bruck's communication pattern to support efficient sparse reconfiguration. The key idea is to reduce propagation and transmission delay by directly connecting immediate communication partners and preserve efficient reachability to future peers through connected \textit{subrings}. As a result, optical links can be \textit{reused} across multiple subsequent steps, allowing the benefit of reconfiguration to \textit{amortize} beyond a single step. Our evaluation shows that \name reduces All-to-All completion time by typically \(3\times\) to \(10\times\) over static baselines and is effective even with millisecond-scale reconfiguration delays; for AllReduce up to \(1.5\times\) against existing reconfiguration strategies, and by $1.5\times$ to $6.6\times$ against the \ring algorithm on low to moderate-sized workloads.
\end{abstract}

\maketitle

\section{Introduction}
The rapid growth of deep learning models and datasets has made both training and inference increasingly dependent on large-scale distributed infrastructure. To meet these demands, the four leading US hyperscalers have reportedly announced plans to invest \$650~billion in AI infrastructure in 2026 alone~\cite{reuters2026aiinfra}. A considerable share will go into GPU- and TPU-powered datacenters, where performance depends critically on collective communication among accelerators, particularly AllReduce and All-to-All~\cite{Khani2021SiP}.

The design of datacenter interconnect fabrics therefore plays a central role in the scalability and efficiency of large-scale ML systems~\cite{Qian2024Alibaba, Jouppi2023TPUv4, Ding2025Rails}. Whereas electrically switched networks are power-intensive and may lead to performance bottlenecks, optical networks have emerged as a promising opportunity to reconfigure network topology at runtime to improve both performance and efficiency~\cite{Ding2025Rails, Kumar2024Case}. Optical circuit switches (OCSs) establish optical paths between pairs of GPUs by steering light from input to output ports~\cite{Ding2025Rails}. As these paths can be reconfigured, the network topology can be aligned with the communication traffic of the active workload~\cite{Avin2025Rev, Zerwas2023Duo, Avin2019DAN, Ding2025Rails, Vamsi2025Bend}. Collective communication is particularly compatible with optical reconfigurable networks (ORNs): it largely follows one-to-one data exchanges, the communication pattern for which the topology can be optimized is known in advance, and collectives proceed in synchronized phases, between which the network can be reconfigured~\cite{Ding2025Rails, Weiyang2023TopoOpt, Kumar2024Case, Gangidi2024}. This motivates low-degree topologies such as rings, which preserve end-to-end connectivity among all nodes at degree two using \(2n\) optical ports~\cite{ Ding2025Rails}. 

\begin{figure}[b]
    \centering
    \includegraphics[width=\columnwidth]{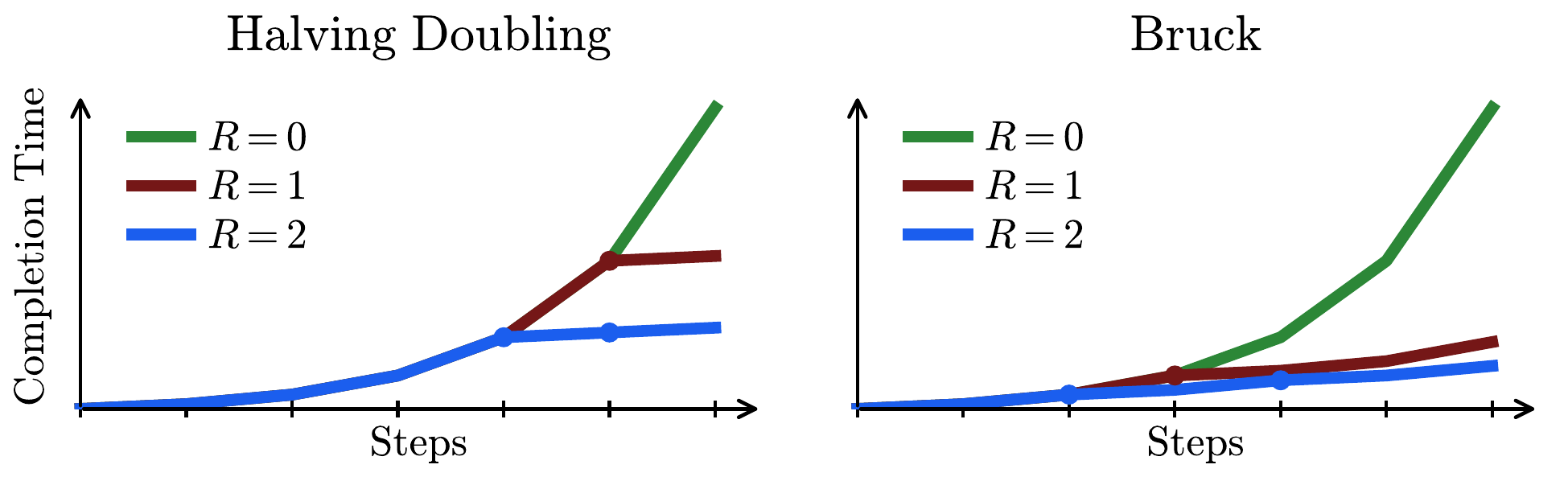}
    \caption{Cumulative AllReduce communication cost of \bruck compared to \hd for $n=64$ with $R=0,1,2$ reconfigurations (reconf. delay is not considered).}
  \label{fig:intro}
\end{figure}

Prior work on collective communication for ORNs explored different directions: one approach, typically considering no or negligibly low reconfiguration delay~\cite{Vamsi2023Mars, Amir2024}, concludes that every-step topology adaptations are beneficial; commonly derived from Birkhoff--von Neumann (BvN) decompositions~\cite{birkhoff1946three} of aggregate traffic matrices~\cite{Khani2021SiP, Vamsi2025Bend, Kumar2024Case}. Second, systems such as \textsc{TopoOpt} \cite{Weiyang2023TopoOpt} that reconfigure once before the workload assume that adapting the network during collective operations incurs prohibitive overhead due to high reconfiguration delay. A third line explicitly trades off reconfiguration delay against completion time~\cite{Vamsi2025Bend, Rahman2026Harvest, hammer2025}.

\noindent However, existing reconfigurable strategies are often constrained by the underlying communication algorithm. Figure~\ref{fig:intro} visualizes this for 64 nodes by comparing the completion time of state-of-the-art algorithms, \halving(\hd)~\cite{Thakur2005} and \bruck~\cite{Bruck94}. For \textsc{HD} all schedules have identical cost until reconfigurations start, whereas \bruck reconfigures earlier and \textit{reuses} reconfigurable links to form \textit{subrings} which reduce costs for subsequent steps significantly~\cite{Rahman2026Harvest}.

This paper is motivated by the observation that existing reconfiguration algorithms are myopic and miss an important optimization opportunity: optical links can be \textit{reused} by providing a \name between communicating peers of current and later steps. Thereby, the cost of a reconfiguration can be \textit{amortized} over all future steps.

We propose \name, a reconfiguration strategy that constructs \textit{subrings}, consisting of active communication pairs, which preserve reachability to peers in subsequent steps. Based on \bruck's algorithm, we derive optimal reconfiguration schedules for AllReduce and All-to-All. As a result, once the communication of a reconfigured step is completed, the same topology can be \textit{reused} across future steps without further reconfiguration, while peers are placed closer together in a \textit{subring}. This enables high reconfiguration overheads to be \textit{amortized} across multiple steps, increasing its benefit.

In our evaluation, \name reduces AllReduce completion time consistently compared to existing reconfiguration strategies with speedups of up to \(1.5\times\) and outperforms the bandwidth-optimal \ring algorithm by $1.5\times$ to $6.6\times$ on low to moderate-sized workloads. \name improves All-to-All completion time over static solutions by typically up to \(10\times\) at low reconfiguration delays, and even by up to \(5\times\) when reconfiguration delays are in the milliseconds.

Our key contributions are:
\begin{itemize}
    \item We introduce \name, a novel approach that leverages \bruck's algorithm to realize AllReduce and All-to-All by \textit{reusing} optical links, thereby enabling sparse, infrequent or even single reconfigurations (Section~\ref{sec:bridge}).
    \item Based on \bruck, we characterize the optimal reconfiguration schedules for All-to-All, Reduce-Scatter, and AllGather for any fixed number of reconfigurations (Sections~\ref{subsec:alltoall}--\ref{subsec:allgather}) and extend to networks with fewer than $2n$ optical ports (Section~\ref{sec:bridge-lessocs}).
    \item We conduct large-scale simulations in Astra-Sim~\cite{Won2023Astrasim} using ns-3~\cite{ns3-simulator} as the network backend, comparing \name for AllReduce (Section~\ref{sec:ev:reduce}) and All-to-All (Section~\ref{sec:ev:alltoall}) against existing reconfiguration strategies. Our evaluation spans message sizes from \(1\,\mathrm{KB}\) to \(256\,\mathrm{MB}\), realistic reconfiguration delays from \(1\,\mu\mathrm{s}\) to \(5\,\mathrm{ms}\), and a wide range of network sizes and per-hop delays.
\end{itemize}

\section{Motivation}
\label{sec:motivation}
Collective communication primitives specify communication objectives among a group of GPUs, such as AllReduce, Reduce-Scatter, AllGather and All-to-All~\cite{Gangidi2024,Thakur2005}. Collective algorithms realize these primitives by following a predefined communication pattern~\cite{Thakur2005, juerss2026}. They are typically organized as a sequence of \textit{communication steps}, in each of which a GPU sends data to exactly one other GPU~\cite{Thakur2005, Daniele2024Swing}. In AllReduce, each node contributes data to a global reduction whose result is returned to all nodes, often via a Reduce-Scatter followed by an AllGather~\cite{Thakur2005}. In All-to-All, each GPU sends a distinct message to every other GPU. These collectives can be implemented with a variety of algorithms~\cite{Bruck94,Thakur2005,juerss2026,Daniele2024Swing, traff2025} and place substantial demands on the network, especially All-to-All.

\textbf{Optical Reconfigurable Networks (ORNs)} offer an opportunity to address this bandwidth overhead by adapting connectivity at runtime~\cite{Jouppi2023TPUv4, Khani2021SiP, Vamsi2025Bend, Amir2024}. Unlike packet-switched electrical fabrics, however, an OCS provides only one-to-one circuits at any moment of time~\cite{Ding2025Rails, Mellette2017RotorNet}, which naturally leads to low-degree topologies. Ring-like fabrics are therefore a practical design for OCS-based distributed ML systems~\cite{Ding2025Rails, Kumar2024Case, Vamsi2025Bend}. Recent work exploits the rail structure of scale-out systems---fixed groups of links or nodes communicate within the same parallel communication lane---by replacing packet switches within a rail with OCSs~\cite{Ding2025Rails}. This realizes direct-connect rings that preserve connectivity at low node degree while enabling reconfiguration across communication phases. 

A common choice to implement All-to-All is to complete the operation in \(n-1\) parallel point-to-point exchanges offering no opportunity to reconfigure the network to reduce particularly costly traffic~\cite{Thakur2005, Qin2025}. For AllReduce, \halving (\hd) is a common latency-optimal implementation where in step $k$ each node $u$ communicates to $v = u \oplus 2^k$ (recursively halves/doubles distance) for $\log_2 n$ steps, whereas \ring is bandwidth-optimal~\cite{Hu2025NCCL} and performs $n-1$ clockwise single-block transmissions~\cite{Thakur2005}. \textbf{\bruck} provides a promising All-to-All and also AllReduce algorithm for ORNs, where each node, given a step $k$, communicates to its peer at offset \(2^k\). This completes All-to-All or AllReduce in \(\log_2 n\) steps. On static fabrics, \bruck matches the cost of \hd, but replaces its pairwise exchanges with a cyclic communication pattern which naturally forms connected \textit{subrings}. To assess the \textbf{performance} of collective communication algorithms, the Hockney $\alpha-\beta$ cost model~\cite{hockney} is commonly used. In line with related work~\cite{Won2023Astrasim, Rahman2026Harvest, hammer2025, juerss2026, Juerss2026RevisitingBruck, Juerss2027Sigmetrics, Aashaka2023TACCL}, we use an extension of this cost model to account for a congestion factor $c$ (overlapping flows per link), decompose network latency into a per-hop latency \(\alpha_h\) (propagation delay and per-hop message processing)~\cite{hoefler2010toward} and a per-step latency \(\alpha_s\). Further, network reconfiguration overheads are modeled with $\delta$ as follows:
\begin{equation*}
    T(m,A) = \sigma(A) \cdot \alpha_s + \sum_{k=0}^{\sigma(A)-1} (h_k \cdot \alpha_h) + \sum_{k=0}^{\sigma(A)-1} (m_{k} \cdot c_{k} \cdot \beta) + R\cdot \delta
\end{equation*}
where, in each step $k$ of total $\sigma(A)$ steps, the algorithm $A$ incurs a startup latency $\alpha_s$ (e.g., data preparation), per-hop delay $\alpha_h$ for each hop $h_k$ required to reach its destination, a transmission delay of $\beta \cdot m_k \cdot c_k$ ($m_k$ is the chunk size transmitted in step $k$, $\beta$ the network cost per byte --- inverse of bandwidth $b$) and a reconfiguration overhead $\delta$ for the number of reconfigurations $R$. $\delta$ accounts for the aggregate overhead incurred between consecutive communication steps due to reconfiguration, including switch reconfiguration and topology adjustment, link bring-up, control-plane operations, and node synchronization~\cite{Ding2025Rails, Porter2013}. We omit computation costs, as these are similar among collective algorithms~\cite{Thakur2005}.

\begin{figure}[t]
    \centering
    \includegraphics[width=\columnwidth]{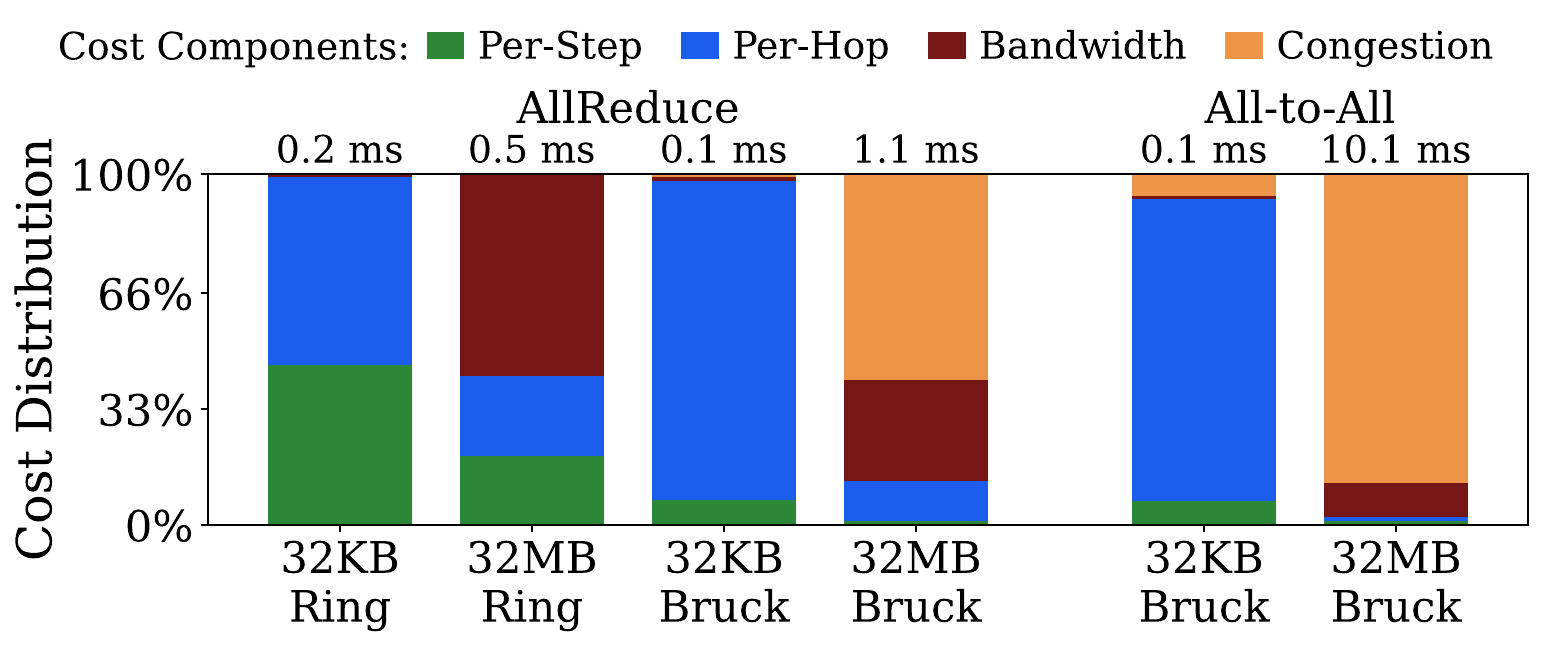}
    \caption{Cost distribution and completion time of state-of-the-art algorithms for AllReduce and All-to-All. The opportunity for improvement through ORNs is greater for All-to-All (optimize per-hop delay and congestion), whereas for AllReduce it is more limited for large workloads (only per-hop delay can be optimized as \ring minimizes congestion).}
  \label{fig:costdistr}
  \vspace{-15px}
\end{figure}

Although recent \textbf{static approaches} for rings and tori use topology-aware scheduling or mirroring to reduce congestion and path length~\cite{Daniele2024Swing, juerss2026}, they cannot remove the fundamental costs imposed by the topology itself. Figure~\ref{fig:costdistr} shows the completion time distribution of \ring and \bruck in Astra-Sim~\cite{Won2023Astrasim} for AllReduce and All-to-All on a static ring topology. ORNs directly connect peers, reducing the number of hops and congestion on each link. For All-to-All, per-hop delay dominates at small workloads and congestion at large ones, making direct connections through OCSs desirable. For AllReduce, \bruck shows a similar trend, but the bandwidth-optimal \ring highlights a key limitation: for large workloads, completion time is dominated by unavoidable transmission cost ($m\cdot\beta$) rather than congestion overhead. Consequently, the potential of reconfiguration is larger for All-to-All, while for AllReduce it is strongest for small workloads and more constrained for large ones.

\textbf{Prior work} on leveraging ORNs for collectives follows different directions: the topology is reconfigured greedily at every step when reconfiguration delay is assumed negligible~\cite{Amir2024, Porter2013, Vamsi2023Mars}, or it is configured once before the collective begins and then kept static throughout execution when reconfiguration delay is high~\cite{Weiyang2023TopoOpt, Jouppi2023TPUv4}. Beyond these extremes, recent work trades off reconfiguration overhead with communication cost savings~\cite{hammer2025, Vamsi2025Bend}, e.g., on dynamic reconfiguration for \hd~\cite{Vamsi2025Bend}, which we refer to as \rehd. However, \rehd remains fundamentally limited because each reconfiguration primarily benefits only the current step. The \rehd reconfiguration strategy initially operates on a ring until the first reconfiguration is performed. Then, each node can only reach its current peer and these connections are not useful in later steps. As a result, the topology must be reconfigured again to proceed~\cite{Vamsi2025Bend}. Each reconfiguration therefore benefits only a single step and must justify the delay of each reconfiguration. According to our evaluation (Section~\ref{sec:eva}), the gain from one shortcut is often too small to amortize the cost of reconfiguration. Concurrently with this work, Harvest~\cite{Rahman2026Harvest} observes that cyclic patterns preserve reachability to subsequent peers which it uses to reduce the topology search space. For hybrid topologies and All-to-All communication, in-collective reconfiguration using \bruck with a single optical matching augmenting an electrical ring has been studied~\cite{Juerss2027Sigmetrics}; \name instead uses two optical ports per node, forming connected subrings in which current peers communicate at minimal distance while future peers remain reachable. Therefore, how to sparsely reconfigure and \textit{reuse} optical links depending on the collective workload remains unexplored.

The \textbf{main challenge} in optical reconfigurable networks is that the communication-optimal strategy---directly reconnecting active pairs in every step---quickly becomes impractical when reconfiguration delay is non-negligible~\cite{Vamsi2025Bend, Mellette2017RotorNet, Jouppi2023TPUv4}. Reconfiguration schedules must therefore be self-sustaining, in the sense that instead of disrupting the remaining communication pattern, they should at minimum preserve and ideally improve it. Therefore, we posit that a single, efficient reconfiguration needs to satisfy three conditions:

\begin{itemize}[leftmargin=*]
   \item[] \textbf{Condition~1:} \emph{Minimize communication costs in the current step}
    \item[] \textbf{Condition~2:} \emph{Preserve reachability to peers of subsequent steps}
    \item[] \textbf{Condition~3:} \emph{Reduce costs for subsequent steps by established, \textit{reusable} reconfigurable links without further reconfiguration}
\end{itemize}

\noindent To minimize communication cost in the current step, a necessary condition is that reconfigurations directly connect active communication pairs~\cite{Vamsi2025Bend, Rahman2026Harvest, Weiyang2023TopoOpt}. The second condition addresses the key requirement for sparse reconfiguration: beyond optimizing the current step, a reconfiguration must preserve sufficient reachability for the remaining communication pattern so that the collective can continue without immediate further reconfiguration~\cite{Rahman2026Harvest}. Condition~3 strengthens this requirement further. Preserving reachability alone allows the collective to proceed, but an effective and self-sustained reconfiguration should also reduce the cost of subsequent steps by \textit{reusing} established reconfigurable links useful beyond the current exchange. This enables the cost of a single reconfiguration to be \textit{amortized} across multiple subsequent steps, thereby substantially reducing communication costs and strengthening the case for reconfiguration. More precisely, as optical ports are limited in number, ring-like topologies are the natural design structure~\cite{Ding2025Rails}. Satisfying Conditions~1--3 requires reconfiguration schedules that form connected \textbf{Subrings}: they must directly connect the active communication pairs, while at the same time creating a \name to future peers through \textit{reusable} reconfigurable links to reduce costs of later steps.

\section{\name}
\label{sec:bridge}
In this section, we introduce \name: a reconfiguration strategy based on \bruck's communication pattern. First, we establish in Section~\ref{sec:bridge-bruck} its reconfiguration strategy for All-to-All and AllReduce. We then show in Section~\ref{sec:bridge-subring} how \bruck can be leveraged to address the fundamental challenges of reconfiguration scheduling. Building on this, we derive and prove for \bruck the optimal reconfiguration schedules of \name for All-to-All (Section~\ref{subsec:alltoall}) and Reduce-Scatter (Section~\ref{subsec:reducescatter}), which translate to optimal schedules for AllGather (Section~\ref{subsec:allgather}). In Section~\ref{sec:optimal-number-reconfigurations}, we outline how to determine the optimal number of reconfigurations. Finally, Section~\ref{sec:bridge-lessocs} extends the model to settings with fewer than \(2n\) optical ports.

\subsection{Approach: Reusable Subrings with \bruck}
\label{sec:bridge-bruck}

We consider a scale-up network of \(n\) nodes (power-of-two) interconnected by a programmable OCS fabric initially arranged as a ring~\cite{Ding2025Rails, Kumar2024Case}. The fabric provides \(2n\) optical ports, so that each node maintains one incoming and one outgoing optical connection and thus communicates with exactly one peer at a time~\cite{Ding2025Rails}.

Our approach builds on \bruck's communication pattern~\cite{Bruck94,traff2025}, which is particularly well suited to reconfigurable networks. For All-to-All and Reduce-Scatter under a single-port communication model~\cite{Daniele2024Swing, traff2025, Thakur2005}, \bruck completes in $s := \lceil \log_2 n \rceil$ steps. In step \(k \in \{0,\dots,s-1\}\), each node \(u\) communicates with node $u + 2^k \bmod n$, that is, with a peer at ring distance \(2^k\). For All-to-All, every step transmits the same amount of data, namely \(\frac{m}{2}\). For Reduce-Scatter, the same communication pattern can be used, while the transmitted data is determined by the standard block propagation algorithm~\cite{juerss2026, Daniele2024Swing, Thakur2005}. On static fabrics, \bruck exhibits the same number of steps and the same aggregate hop count, congestion, and transmitted data as \hd, but follows a cyclic rather than pairwise communication pattern. For arbitrary \(n\), the algorithm progresses identically except for the last step: if \(2^{s-1} < n < 2^s\), then each node sends only \((m/n)(n-2^{s-1})\) data in step \(s-1\). In multiport networks with $p$ local ports per node, several independent communication offsets can be scheduled in parallel within one step, effectively collapsing \bruck to $\log_{p+1} n$ steps.

\begin{figure}[t]
    \centering
    \includegraphics[width=0.8\columnwidth]{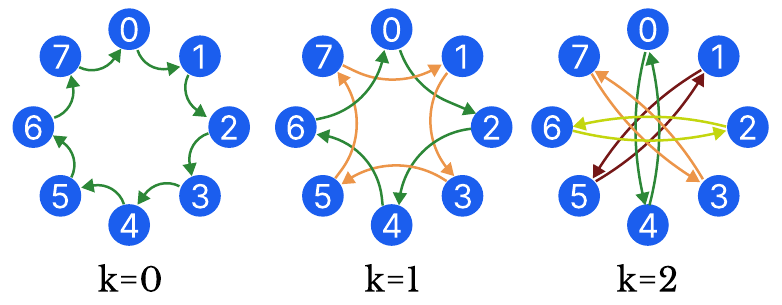}
    \caption{\bruck's communication pattern for 8 nodes separated in 3 steps. Reconfigurations in each step would result in the displayed topology. The OCS is omitted for clarity. }
  \label{fig:bruck}
  \vspace{-10px}
\end{figure}

Figure~\ref{fig:bruck} visualizes the key property of \bruck's communication pattern across consecutive steps: \emph{transitivity}. If node \(u\) communicates with node \(v\) in step \(k\), and node \(v\) communicates with node \(w\) in the same step, then node \(u\) communicates with node \(w\) in step \(k+1\), since $2^{k+1} = 2^k + 2^k$. Thus, placing peers of the current step close together also keeps future peers close. Furthermore, these peer connections form a closed subring of size \(\frac{n}{2^k}\), containing both all active peers and all future peers reachable from them. This directly addresses the three requirements from Section~\ref{sec:motivation}: it enables minimal cost in the current step, preserves reachability for subsequent steps, and reduces their cost as well through \textit{reusable} links.

\begin{algorithm}[b]
  \caption{\name: Reusable OCS Subrings with \bruck}
  \label{alg:bridge}
  \begin{algorithmic}[1]
    \REQUIRE \(n = 2^s\) nodes, reconfiguration schedule \(\mathbf{x}=(x^0,\dots,x^{s-1})\)
    \STATE \(s \gets \log_2 n\)
    \FOR{each step \(k \in \{0,\dots,s-1\}\)}
        \IF{\(x^k = 1\)}
            \STATE Construct subrings \(S_i^{(k)}\) for offset \(2^k\)
            \FOR{each node \(u\)}
                \STATE Set optical link \((u,\; u + 2^k \bmod n)\)
            \ENDFOR
        \ELSE
            \STATE Reuse the current topology
        \ENDIF
        \STATE Perform step \(k\) of \bruck on the current topology
    \ENDFOR
  \end{algorithmic}
\end{algorithm}

To formalize this process, let $\mathbf{x} = (x^0,\dots,x^{s-1}) \in \{0,1\}^s$ denote a reconfiguration schedule, where \(x^k = 1\) indicates that the topology is reconfigured immediately before step \(k\), and \(x^k = 0\) means that the topology from the previous step is \textit{reused}. The total number of reconfigurations is then $R := \sum_{k=0}^{s-1} x^k$ and a schedule with $R$ reconfigurations is denoted as $\mathbf{x}_R$. Algorithm~\ref{alg:bridge} summarizes \name, where in each communication step \(k\), the schedule \(\mathbf{x}\) determines whether the OCS is reconfigured to the subring associated with offset \(2^k\) or whether the previously established topology is \textit{reused}. For a given reconfiguration schedule \(\mathbf{x}\) and a collective algorithm $A$, we further define the stepwise improvement over the static fabric as $\Delta(\mathbf{x},A) = \bigl(\Delta_0(\mathbf{x},A),\dots,\Delta_{s-1}(\mathbf{x},A)\bigr)$
where \(\Delta_k(\mathbf{x},A)\) denotes the reduction in communication time of step \(k\) relative to the static algorithm $A$, while \(R\cdot\delta\) captures its overhead.

\subsection{Minimal Connected Subrings}
\label{sec:bridge-subring}
To ensure that reconfiguration preserves both locality and connectivity, we next formalize the formation of minimal connected subrings which can be directly derived from \bruck's communication pattern. Such subrings contain exactly the nodes that must remain reachable after a reconfiguration step. Formally, for each subring \(i \in \{0,\dots,2^k-1\}\) in step $k$, we define \(S_i^{(k)}\) as
\[
    S_i^{(k)} := \{\, u \in \{0,\dots,n-1\} \mid u \equiv i \pmod{2^k} \,\}.
\]
Thus, the network is partitioned into \(2^k\) subrings of size \(\frac{n}{2^k}\), and each node \(u\) is connected to \(u+2^k \bmod n\) (to which it sends data) and \(u-2^k \bmod n\) (from which it receives data). Hence, the current communication peer \(u+2^k \bmod n\) is directly adjacent, while future \bruck peers remain reachable within the same subring.

\begin{lemma}
  For step \(k\), the subring \(S_i^{(k)}\) is minimal: it contains exactly the current peer, all future peers, and peers of peers reachable under \bruck from step \(k\) onward.  
\end{lemma}

\paragraph{Proof.}
In step \(k\), node \(u\) communicates with \(u+2^k\), so its current peer lies in the same residue class modulo \(2^k\). In every subsequent step \(j>k\), the communication offset is \(2^j = 2^{j-k}\cdot 2^k\), i.e., again a multiple of \(2^k\). Therefore, every future peer of \(u\), and recursively every peer of such a peer, remains in the same residue class modulo \(2^k\), namely in \(S_i^{(k)}\). Since this residue class contains exactly \(\frac{n}{2^k}\) nodes, and every node in it is reachable by repeated communication steps of size \(2^k,2^{k+1},\dots\), the subring contains exactly the nodes needed to preserve current and future reachability, and no more. \qed

This shows that \name forms minimal subrings in which reconfigurable links can be \textit{reused} to reach future peers. Consequently, for a reconfiguration schedule \(\mathbf{x}\), the cost of a topology reconfiguration can be \textit{amortized} across multiple steps: whenever \(x^k = 1\), the reconfiguration at step \(k\) reduces not only the cost of that step, but also that of subsequent ones, i.e., \(\Delta_i(\mathbf{x}, \bruck) > 0\) for \(i > k\).

\begin{figure}[b]
    \centering
    \includegraphics[width=\columnwidth]{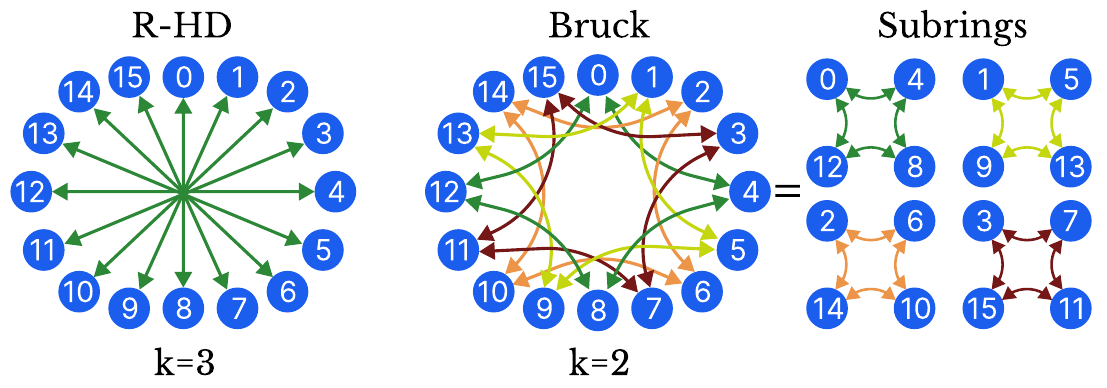}
    \caption{Network topologies for \(n=16\) and \(R=1\): \rehd reconfigured at step \(k=3\) (left), and \name at step \(k=2\) (middle/right). For clarity, the OCS connecting the nodes is omitted. Subrings (right) are grouped from \bruck at $k=2$.}
  \label{fig:subrings}
  \vspace{-15px}
\end{figure}

\noindent Figure~\ref{fig:subrings} illustrates the subring formation for \(R=1\) and \(n=16\). In this example, \bruck is reconfigured at step \(k=2\), leading to \(2^2=4\) subrings of size \(4\). As a result, each node, for example node \(0\), reaches its communication peer node $4$ in a single hop at step \(k=2\), while requiring only two hops at step \(k=3\) to node $8$. By contrast, \rehd benefits from reconfiguration only at step \(k=3\), since directly connecting peers earlier does not preserve the connectivity required for subsequent steps. Thus, for \rehd only \(\Delta_3(\mathbf{x}_{R=1},\rehd) > 0\), whereas for \name both \(\Delta_2(\mathbf{x}_{R=1},\name) > 0\) and \(\Delta_3(\mathbf{x}_{R=1},\name) > 0\)  leading to \(\sum_k \Delta_k(\mathbf{x}_R^\star,\name) \geq \sum_k \Delta_k(\mathbf{x}_R^\star,\rehd)\) for all \(R \in \{0,\dots,s-1\}\). As we show in Section~\ref{sec:eva}, this enables \name to outperform \rehd by up to \(1.5\times\) in completion time.

\subsection{All-to-All: Periodic Reconfigurations}
\label{subsec:alltoall}

To minimize communication cost for \bruck, we first prove that periodic reconfigurations are optimal (Theorem~\ref{th:periodic}) and then determine in Section~\ref{sec:optimal-number-reconfigurations} how many reconfigurations are worthwhile. We characterize the reconfiguration space of the algorithm across \(s=\log_2 n\) steps, where the network may reconfigure at each step. We first find optimal reconfiguration schedules to reduce All-to-All completion time $C_{A2A}$ with \bruck's algorithm, where each node sends per step the same amount of data, $m_k = \frac{m}{2}$. As congestion and communication distance in \bruck are identical $(c_k = h_k)$, the cost contribution of step \(k\) is (total costs are in \(\Theta(n)\)):
\[
\alpha_s + \alpha_h h_k + \beta m_k c_k
=
\alpha_s + (\alpha_h+\beta\frac{m}{2}) h_k.
\]

\noindent Consider a schedule with exactly \(R\) reconfigurations, for which the lengths of segments sum up to $s = \sum_{j=1}^{R+1} r_j$. Each segment captures the progression of \bruck between topology reconfigurations. Within each segment, the communication distance in \bruck starts at \(1\) and doubles at every step, so a segment of length \(r\) has cost
\[
C_{\mathrm{seg}}(r)
=
\sum_{i=0}^{r-1}(\alpha_s + (\alpha_h+\beta\frac{m}{2})2^i)
=
r\alpha_s + \gamma(2^r-1),
\]
where $\gamma := \alpha_h+\beta\frac{m}{2}$. Hence the total cost of a schedule is
\[
C_{A2A}(r_1,\dots,r_{R+1})
=
\sum_{j=1}^{R+1} C_{\mathrm{seg}}(r_j) + R\delta
=
s\alpha_s + \gamma\sum_{j=1}^{R+1}(2^{r_j}-1) + R\delta.
\]
Since \(s\alpha_s\), \(\gamma\), and \(R\delta\) are fixed for a given \(R\), minimizing the total cost is equivalent to minimizing $\sum_{j=1}^{R+1} 2^{r_j}$ subject to $\sum_{j=1}^{R+1} r_j=s$.

\begin{lemma}\label{thm:alltoall-schedules}
For fixed \(R\), every optimal All-to-All schedule has segment lengths that differ by at most one. We refer to such schedules as \emph{periodic}.
\end{lemma}

\paragraph{Proof by Contradiction.}
Define the marginal cost of extending a segment from length \(r-1\) to length \(r\) as
\[
D(r):=C_{\mathrm{seg}}(r)-C_{\mathrm{seg}}(r-1)=\alpha_s+\gamma\,2^{r-1}.
\]
Since \(2^{r-1}\) is strictly increasing, \(D(r)\) is strictly increasing in \(r\). Segments of the same size incur the same cost. Assume an optimal schedule contains two segment lengths \(r_a\) and \(r_b\) with $r_a \ge r_b+2$. In this case, we can move one step from the longer segment to the shorter one, i.e., $r_a' = r_a-1$ and $r_b' = r_b+1$.
The total number of steps remains unchanged. The corresponding change in cost is:
\[
\bigl(C_{\mathrm{seg}}(r_a)+C_{\mathrm{seg}}(r_b)\bigr)
-
\bigl(C_{\mathrm{seg}}(r_a-1)+C_{\mathrm{seg}}(r_b+1)\bigr)
=
D(r_a)-D(r_b+1).
\]
Because \(r_a \ge r_b+2\), we have \(r_a > r_b+1\), and since \(D(r)\) is strictly increasing $D(r_a)>D(r_b+1)$. Therefore
\[
C_{\mathrm{seg}}(r_a)+C_{\mathrm{seg}}(r_b)
>
C_{\mathrm{seg}}(r_a-1)+C_{\mathrm{seg}}(r_b+1),
\]
which contradicts optimality. Hence, no optimal schedule can contain two segments whose lengths differ by at least two, since they could be rearranged to balanced segment sizes. Therefore, in every optimum, $|r_a-r_b|\le 1$ for all $a,b$.
\hfill\(\square\)

\noindent We call such schedules \emph{periodic}, where reconfiguration intervals are spaced as evenly as the discrete communication steps permit. Let \(q:=\lfloor s/(R+1)\rfloor\) and \(t:=s\bmod(R+1)\). By Lemma~\ref{thm:alltoall-schedules}, every optimal schedule consists of \(t\) segments of length \(q+1\) and \(R+1-t\) segments of length \(q\).

\begin{theorem}
    Given \(n=2^s\) nodes and a fixed number \(R\in\{0,\dots,s-1\}\) of reconfigurations, every optimal All-to-All schedule for \bruck is periodic: it partitions the \(s\) communication steps into \(R+1\) segments whose lengths differ by at most one. In particular, if \(R+1\) divides \(s\), reconfigurations occur exactly every \(\frac{s}{R+1}=\frac{\log_2 n}{R+1}\) steps.
    \label{th:periodic}
\end{theorem}
\vspace{-13px}
\noindent The resulting optimal cost is
\[
C_{\mathrm{A2A}}^\star(R)
=
s\alpha_s
+
\gamma\left[
(R+1-t)(2^q-1)
+
t(2^{q+1}-1)
\right]
+
R\delta,
\]
where \(q=\left\lfloor s/(R+1)\right\rfloor\), \(t=s\bmod(R+1)\), and \(\gamma=\alpha_h+\beta\frac{m}{2}\). If \(R+1\) divides \(s\), this simplifies to
\[
C_{\mathrm{A2A}}^\star(R)
=
s\alpha_s
+
(R+1)\gamma(n^{\frac{1}{R+1}}-1)
+
R\delta.
\]

\noindent Thus, for fixed \(R\), periodic reconfigurations reduce the All-to-All communication cost from \(\Theta(n)\) to \(O((R+1)n^{1/(R+1)})\).
\subsection{Reduce-Scatter: Early Reconfigurations}
\label{subsec:reducescatter}

We now search for optimal reconfiguration schedules to complete the AllReduce collective with \bruck. AllReduce is typically decomposed through the Rabenseifner decomposition~\cite{Thakur2005} into a Reduce-Scatter phase followed by an AllGather phase. Specifically for AllGather, we use an optimized version of \bruck's algorithm rearranging the communication pairs~\cite{Rahman2026Harvest, juerss2026, traff2025}. We first identify optimal schedules for Reduce-Scatter where \bruck's algorithm transmits data starting at $m_0=\frac{m}{2}$ which halves in every subsequent step. The cost of step $k \in \{0,\dots,s-1\}$ is
\[
\alpha_s + \alpha_h h_k + \beta m_k c_k
=
\alpha_s + \alpha_h 2^{k} + \beta \frac{m}{2^{k+1}}2^{k}
=
\alpha_s + \alpha_h 2^{k} + \beta\frac{m}{2}.
\]
As established in Theorem~\ref{th:periodic}, the optimal reconfiguration periods to reduce path lengths and latency for \bruck are periodic. In the following, we therefore focus on finding the optimal schedule to minimize transmission delay. Consider one reconfiguration period covering steps $a,a+1,\dots,b$ with $0\le a\le b\le s-1$. The transmission cost of step \(k\) in this period is
\[
\beta m_k c_k
=
\beta \frac{m}{2^{k+1}}2^{k-a}
=
\beta \frac{m}{2^{a+1}},
\]
Since \(\frac{m\beta}{2}\) is constant across all schedules, it can be factored out. Minimizing the sum of all steps in the period yields the ILP:
\[
C^\star_{\mathrm{RS}}(R) = \min \sum_{0\le a\le b\le s-1} \frac{(b-a+1)}{2^a}\,z_{a,b} 
\] subject to $\sum_{\substack{0\le a\le \tau\\ \tau\le b\le s-1}} z_{a,b}=1$ for all $\tau \in \{0,\dots,s-1\} \text{ and } \sum z_{a,b}=R+1$, for $0\le a\le b\le s-1$ where \(z_{a,b}\in\{0,1\}\) indicates whether the interval \([a,b]\) is chosen. Each chosen variable \(z_{a,b}=1\) represents one reconfiguration period covering exactly the steps \(a,\dots,b\). The ILP returns the transmission delay optimal reconfiguration schedule and reveals the optimal placement of reconfigurations, where resetting the distance is most valuable early in the algorithm.

\begin{theorem}\label{thm:reducescatter-ILP}
Given a network of $n$ nodes, $s$ steps and $R$ reconfigurations, \name partitions $s$ in the optimal transmission delay schedule for Reduce-Scatter in \(R+1\) periods minimizing $C^\star_{\mathrm{RS}}(R)$.
\end{theorem}

\noindent For every fixed number \(R\) of reconfigurations, the optimal Reduce-Scatter schedule is different and is given by the ILP above. However, the reconfiguration schedules are independent of network parameters and therefore only need to be computed once. Finally, we observe that for small networks or for a large number of reconfigurations, the transmission-optimal and latency-optimal Reduce-Scatter schedules coincide. Intuitively, as \(R\) increases relative to the total number of steps \(s=\log_2 n\), every period becomes very short, so the space of feasible schedules collapses to partitions consisting of period lengths \(1\) and \(2\). In our evaluation (Section~\ref{sec:eva}), the optimal schedules were computed within milliseconds. Further, the optimal reconfiguration points for Reduce-Scatter occur earlier than the periodic reconfigurations of All-to-All.

\subsection{AllGather: Late Reconfigurations}
\label{subsec:allgather}

AllGather is the reverse communication pattern of Reduce-Scatter~\cite{Thakur2005}. In \bruck's algorithm, the communication distance starts at $h_k = c_k = 2^{s-1-k}$ for $ k=0,\dots,s-1$ while the transmitted data starts at \(m/n\) and doubles in every step~\cite{Bruck94}.
Hence, as in Reduce-Scatter, the transmission cost of one step is constant:
\[
\beta m_k c_k
=
\beta \frac{m}{2^{\,s-k}}2^{s-1-k}
=
\beta\frac{m}{2}.
\]
Since AllGather reverses both the message-size evolution and the communication distances of Reduce-Scatter, the total transmission cost of an AllGather schedule is identical to that of the reversed Reduce-Scatter schedule~\cite{Thakur2005}. Therefore, for any fixed number \(R\) of reconfigurations, the transmission-optimal AllGather schedule is the reverse of the transmission-optimal Reduce-Scatter schedule. Consequently, whereas Reduce-Scatter places reconfigurations early, AllGather places them late. If $(r_1^\star,\dots,r_{R+1}^\star)$ is the transmission-optimal schedule for Reduce-Scatter, $(r_{R+1}^\star,\dots,r_1^\star)$
is optimal for AllGather. Regarding reconfiguration schedules for optimal latency, periods are balanced as in Section~\ref{subsec:alltoall}.

Unlike Reduce-Scatter, subsequent communication partners are not reached naturally after a reset since communication distance is strictly decreasing. Before the collective starts, we construct the \textit{subring} to preserve reachability until the next reconfiguration step $k$, effectively connecting communication peers of step $k-1$. Since the communication distance decreases, all nodes remain reachable until the next reconfiguration, while the cost is minimized at step \(k-1\). The optimization is also reversed compared to Reduce-Scatter: in Reduce-Scatter, we minimize the cost of the immediate next step and let it double thereafter, whereas in AllGather, we halve the cost at each step until it reaches its minimum just before the next reconfiguration. At step \(k\), we reconfigure the topology according to \bruck for the next reconfiguration step, or into a ring if no further reconfiguration occurs. This leads \name to delay AllGather reconfigurations relative to a periodic schedule.
\begin{table}[H]
\vspace{-10px}
\centering
\small
\begin{tabular}{l|cccccc}
    & $k=0$ & $1$ & $2$ & $3$ & $4$ & $5$ \\
\hline
$x_{R=1}(\name_{\text{All-to-All}})$ & 0 & 0 & 0 & 1 & 0 & 0 \\
$x_{R=1}(\name_{\text{Reduce-Scatter}})$ & 0 & 0 & 1 & 0 & 0 & 0 \\
$x_{R=1}(\name_{\text{AllGather}})$ & 0 & 0 & 0 & 0 & 1 & 0 \\
\hline
$x_{R=2}(\name_{\text{All-to-All}})$ & 0 & 0 & 1 & 0 & 1 & 0 \\
$x_{R=2}(\name_{\text{Reduce-Scatter}})$ & 0 & 1 & 0 & 1 & 0 & 0 \\
$x_{R=2}(\name_{\text{AllGather}})$ & 0 & 0 & 0 & 1 & 0 & 1 \\
\end{tabular}
\caption{Reconfiguration schedules for \name with $n=64$ and $R=1 \text{ or } 2$, $x^k = 1$ indicates reconfiguration before step $k$.}
\label{tab:schedules}
\vspace{-15px}
\end{table}

\noindent Table~\ref{tab:schedules} compares the reconfiguration schedules computed by \name for 64 nodes. For All-to-All, the schedules are strictly periodic. For Reduce-Scatter, they shift one step earlier for both $R=1$ and $R=2$, whereas for AllGather they shift correspondingly later.

\subsection{How often are Reconfigurations beneficial?}
\label{sec:optimal-number-reconfigurations}
To determine how often the network should be reconfigured according to previous optimal schedules, the remaining optimization problem is to minimize the total cost over all feasible values of \(R\), with \(0 \leq R \leq s-1\), which is possible to compute in polynomial time. For All-to-All the optimal number of reconfigurations is
\[
R_{\mathrm{A2A}}^\star
=
\arg\min_{0\le R\le s-1} C^\star_{\mathrm{A2A}}(R).
\]
For Reduce-Scatter, the latency-optimal case is identical to All-to-All, so the optimal number of reconfigurations is again obtained by minimizing the
cost for periodic schedules over \(R\). For the transmission-optimal case, however, the schedule is not balanced and the exact optimum is given by the interval formulation of Theorem~\ref{thm:reducescatter-ILP}. The optimal number of reconfigurations is then
\[
R_{\mathrm{RS}}^\star
=
\arg\min_{0\le R\le s-1} C^\star_{\mathrm{RS}}(R).
\]
Since the optimal schedules for Reduce-Scatter diverge between latency- and bandwidth-constrained configurations, the relevant optimum is the minimum completion time across both cases, namely \(R_{\mathrm{RS}}^\star\) for the bandwidth- and \(R_{\mathrm{A2A}}^\star\) for the latency-dominated setting.
\subsection{Networks with fewer than $2n$ OCS ports}
\label{sec:bridge-lessocs}
For settings in which the OCS provides fewer than \(2n\) optical ports, we extend the model to hierarchically connected static blocks that share reconfigurable optical links in order to preserve a connected subring. Let \(z\) denote the number of available optical ports. Then blocks of \(\lceil\frac{2n}{z}\rceil\) consecutive nodes share two optical ports, one in each direction, and together form a hierarchical ring. This abstraction is motivated by hierarchical systems such as Google's TPUv4~\cite{Jouppi2023TPUv4}, where blocks of chips communicate locally over an electrical fabric and access the optical layer only at the block boundary. Within each block, nodes communicate via static electrical links, while only the leftmost and rightmost nodes connect through the OCS to neighboring blocks. Reconfigurable links can still create shortcut paths that improve throughput, while the electrical substrate preserves connectivity to future communication partners without requiring frequent reconfigurations. Compared to our initial assumption of $2n$ OCS ports, reconfigurations with \bruck in the matched block cluster no longer reduce the effective communication distance to one hop, but only to $\lceil\frac{2n}{z}\rceil$. As a result, this strategy is beneficial in sufficiently large networks, where the communication distance in a given step exceeds the size of the matching cluster.

\section{Evaluation}
\label{sec:eva}
In order to complement our analytically derived reconfiguration schedules presented in Section~\ref{sec:bridge}, we assess in this section the performance of \name for All-to-All (Section~\ref{sec:ev:alltoall}) and AllReduce (Section~\ref{sec:ev:reduce}) compared to state-of-the-art reconfigurable algorithms and static baselines for networks between 16 and 256 nodes. To this end, we conduct extensive simulations using Astra-Sim~\cite{Won2023Astrasim} with ns-3~\cite{ns3-simulator} as the network backend. Astra-Sim models the collective communication schedule and synchronization across nodes, while ns-3 captures packet-level routing, link bandwidth and serialization, propagation delay, queuing, congestion, and the configured transport protocol. We report the corresponding simulation parameters explicitly throughout this section. Within Astra-Sim, we implement the \bruck~\cite{Bruck94} collective algorithm together with an optical switch model that enables topology reconfiguration during collective operations. We structure our evaluation for AllReduce and All-to-All around the following research questions:
\begin{itemize}
    \item How do message size, per-hop delay, and network size influence \name's performance gains?
    \item Under what conditions are reconfigurations feasible?
    \item How does \name outperform existing approaches?
\end{itemize}
Our evaluation demonstrates that dynamic reconfigurations with \name during collective operations reduce AllReduce completion time compared to reconfigurable approaches consistently by up to $1.5\times$ and by $1.5\times$ to \(6.6\times\) compared to static baselines on low to moderate-sized workloads. \name improves All-to-All over existing approaches by up to $10.4\times$ compared to the static baseline and by $1.4\times$ even for a reconfiguration delay of $5\,\mathrm{ms}$.

\subsection{Methodology}
For this evaluation, we explore a broad parameter space, using DCQCN~\cite{Zhu2015DCQCN} as a congestion control algorithm, link bandwidths of 200, 400, and 800\,$\mathrm{Gbps}$, and MTU sizes ranging from 1500\,$\mathrm{B}$ (standard Ethernet) to 9000\,$\mathrm{B}$ (jumbo frames), reflecting large-scale cluster deployments and capabilities of modern RNICs/DPUs~\cite{NVIDIABlueField4Datasheet}. Since our conclusions are consistent across this parameter space, we report results for the configuration that is most representative of current high-performance deployments: 800\,$\mathrm{Gbps}$ links and a 4500\,$\mathrm{B}$ MTU. In addition, we evaluate network sizes from 16 to 256 nodes, per-hop delays from 0.1 to $2\,\mu\mathrm{s}$~\cite{Hu2025NCCL}, and message sizes from 1\,$\mathrm{KB}$ to 256\,$\mathrm{MB}$~\cite{Daniele2024Swing, juerss2026}. We assume a per-step latency of 1.7\,$\mu\mathrm{s}$ \cite{Aashaka2023TACCL}, corresponding to typical InfiniBand-class fabrics.

For the reconfiguration delay, we base our settings on representative optical-switch prototypes. Practical optical networks are primarily constrained by two factors: reconfiguration time and the number of available switch ports. In general, technologies supporting larger port counts, and thus larger network scales, tend to incur higher reconfiguration delays~\cite{Khani2021SiP}. For example, Lightmatter's SiP-based switch provides 32 ports with a reconfiguration time of approximately $7\,\mu\mathrm{s}$~\cite{lightmatter2025passage}, while RotorNet's InFocus switch provides 128 ports at approximately $10\,\mu\mathrm{s}$~\cite{Mellette2017RotorNet}. At larger scales, Calient's 3D-MEMS OCS supports up to 320 ports with a reconfiguration time of approximately $15\,\mathrm{ms}$~\cite{calient2022ocsdatasheet}. We therefore evaluate both small-scale networks with microsecond-scale reconfiguration delays and larger networks with delays in the millisecond range. In particular, Figures~\ref{fig:full_all_64} and~\ref{fig:full_reduce_64} use a reconfiguration delay of $10\,\mu\mathrm{s}$, corresponding to a 128-port OCS such as RotorNet, which is sufficient to support 64 nodes under our two-ports-per-node model. AllReduce is typically implemented via the Rabenseifner decomposition~\cite{Thakur2005} as a Reduce-Scatter and AllGather phase. As shown in Section~\ref{subsec:allgather}, AllGather exhibits identical cost under a reversed Reduce-Scatter reconfiguration schedule~\cite{Thakur2005, juerss2026}. Likewise, the latency-optimal AllReduce schedules under \name can be derived directly from All-to-All. For this reason, we report results for All-to-All and Reduce-Scatter, which together also capture latency-optimal AllReduce and AllGather.

\subsection{All-to-All}
\begin{figure}[t]
    \centering
    \begin{subfigure}[t]{0.495\columnwidth}
        \captionsetup{skip=2pt}
        \centering
        \includegraphics[width=\linewidth]{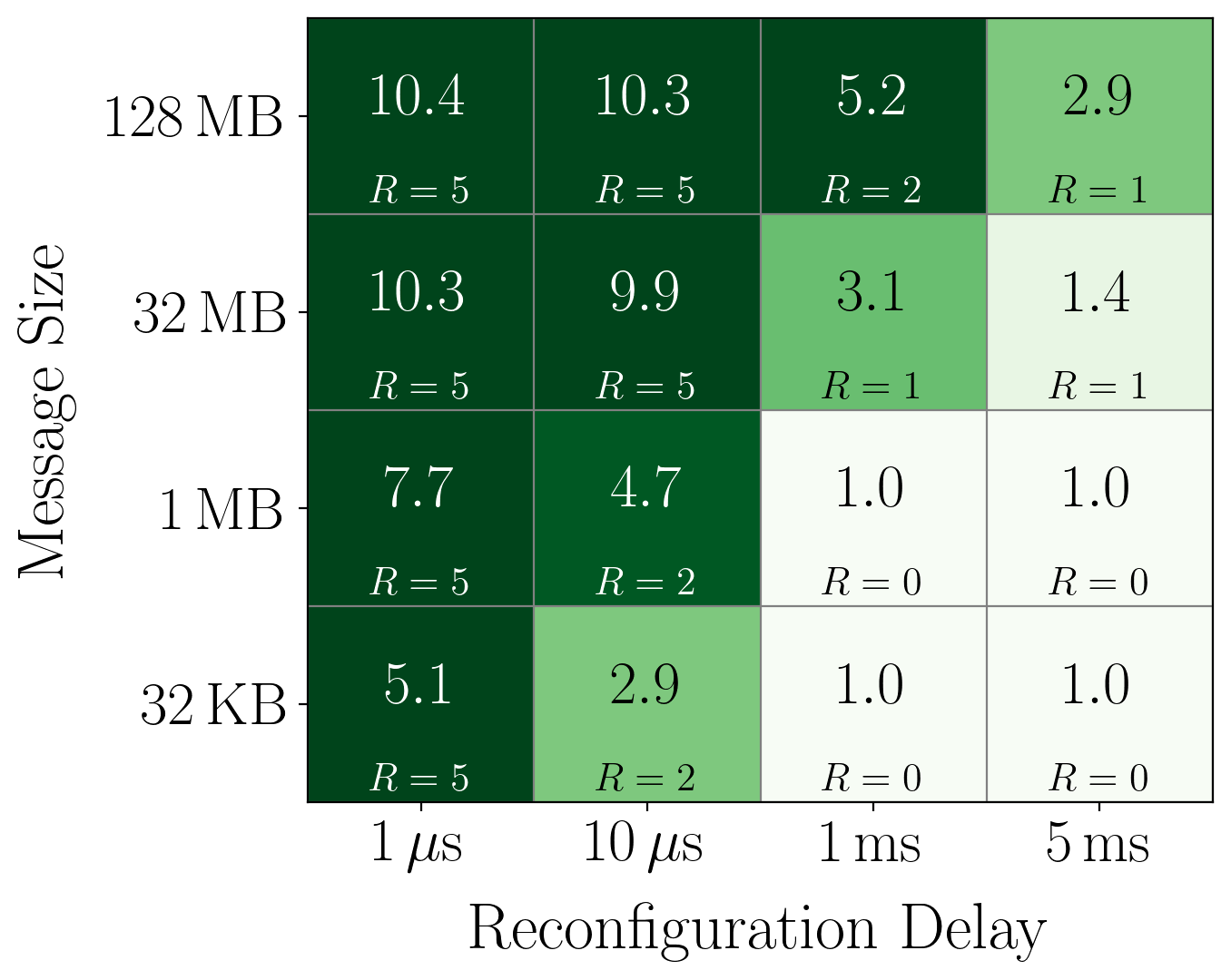}
        \caption{vs. \sbruck}
        \label{fig:matrix_all_msg_static}
    \end{subfigure}
    \hfill
    \begin{subfigure}[t]{0.495\columnwidth}
        \captionsetup{skip=2pt}
        \centering
        \includegraphics[width=\linewidth]{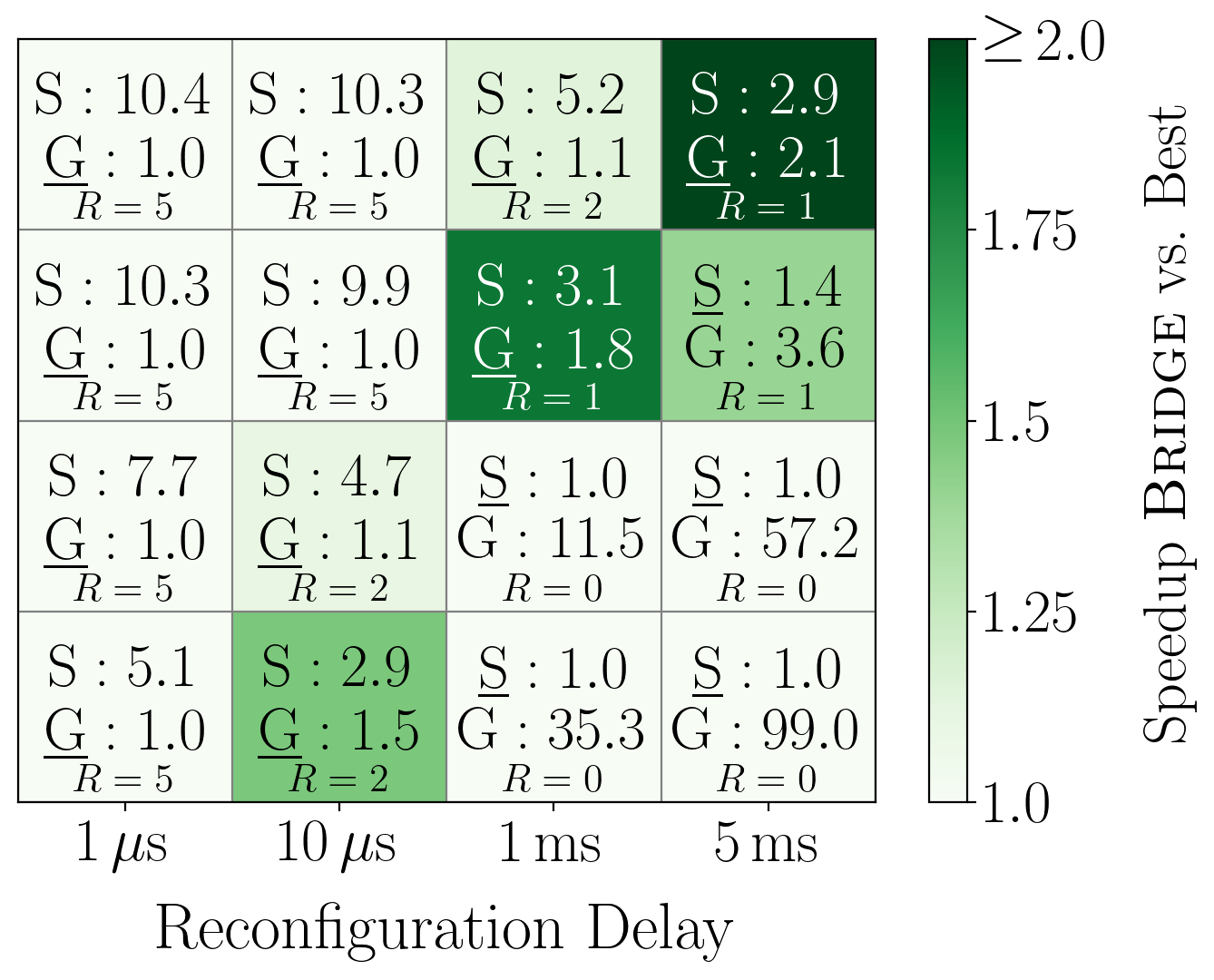}
        \caption{vs. $\min(\sbruck,\gbruck)$}
        \label{fig:matrix_all_msg_all}
    \end{subfigure}
     \caption{Speedup of \name compared to \sbruck and \gbruck for $800\mathrm{Gbps}$ links and $\alpha_h = 1\,\mu\mathrm{s}$ for varying $m$ and $\delta$.}
    \label{fig:matrix_all_msg}
    \vspace{-13px}
\end{figure}

We compare the speedup in completion time achieved by \name for varying message size and reconfiguration delay against static \bruck (\sbruck), which never reconfigures, and greedy \bruck\ (\gbruck), which reconfigures each step based on a BvN decomposition (Figure~\ref{fig:matrix_all_msg}). For varying \textbf{message size}, \name achieves speedups compared to \sbruck of up to \(10.4\times\), with gains generally increasing as reconfiguration delay decreases and message size grows (Figure~\ref{fig:matrix_all_msg_static}). Relative to both baselines in Figure~\ref{fig:matrix_all_msg_all} (the better result is underlined), \name performs best when message size and reconfiguration delay increase proportionally: at high reconfiguration delay and small message sizes, \name behaves similarly to \sbruck, whereas at low reconfiguration delay it approaches \gbruck. The largest gains over both ($\frac{\min(\gbruck,\sbruck)}{\name}$) are observed at \(128\,\mathrm{MB}\) with \(\delta = 5\,\mathrm{ms}\), where it achieves \(2.1\times\) speedup.

\begin{figure}[t]
    \centering
    \begin{subfigure}[t]{0.495\columnwidth}
        \captionsetup{skip=2pt}
        \centering
        \includegraphics[width=\linewidth]{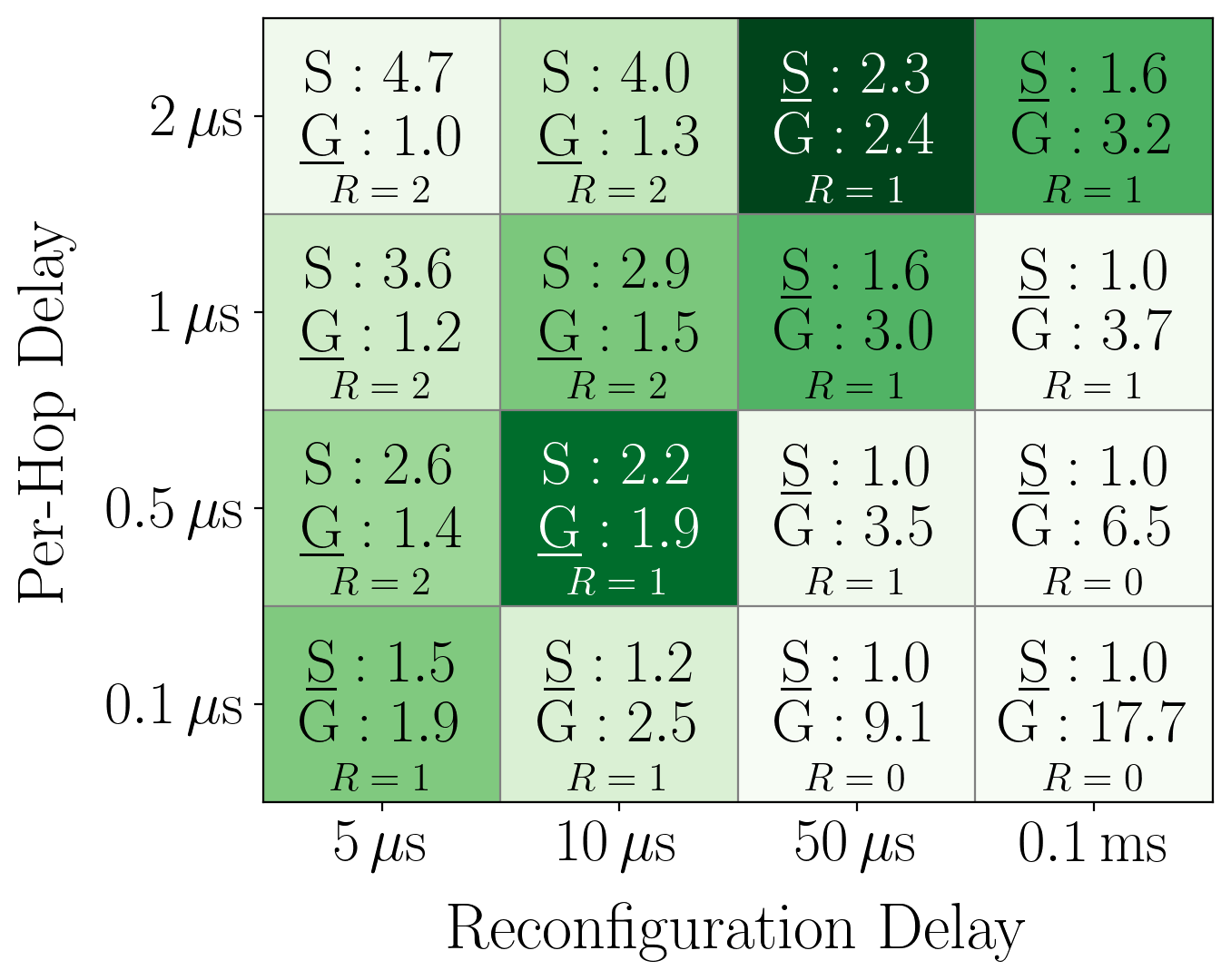}
        \caption{32 KB}
        \label{fig:matrix_all_hop_32kb}
    \end{subfigure}
    \hfill
    \begin{subfigure}[t]{0.495\columnwidth}
        \captionsetup{skip=2pt}
        \centering
        \includegraphics[width=\linewidth]{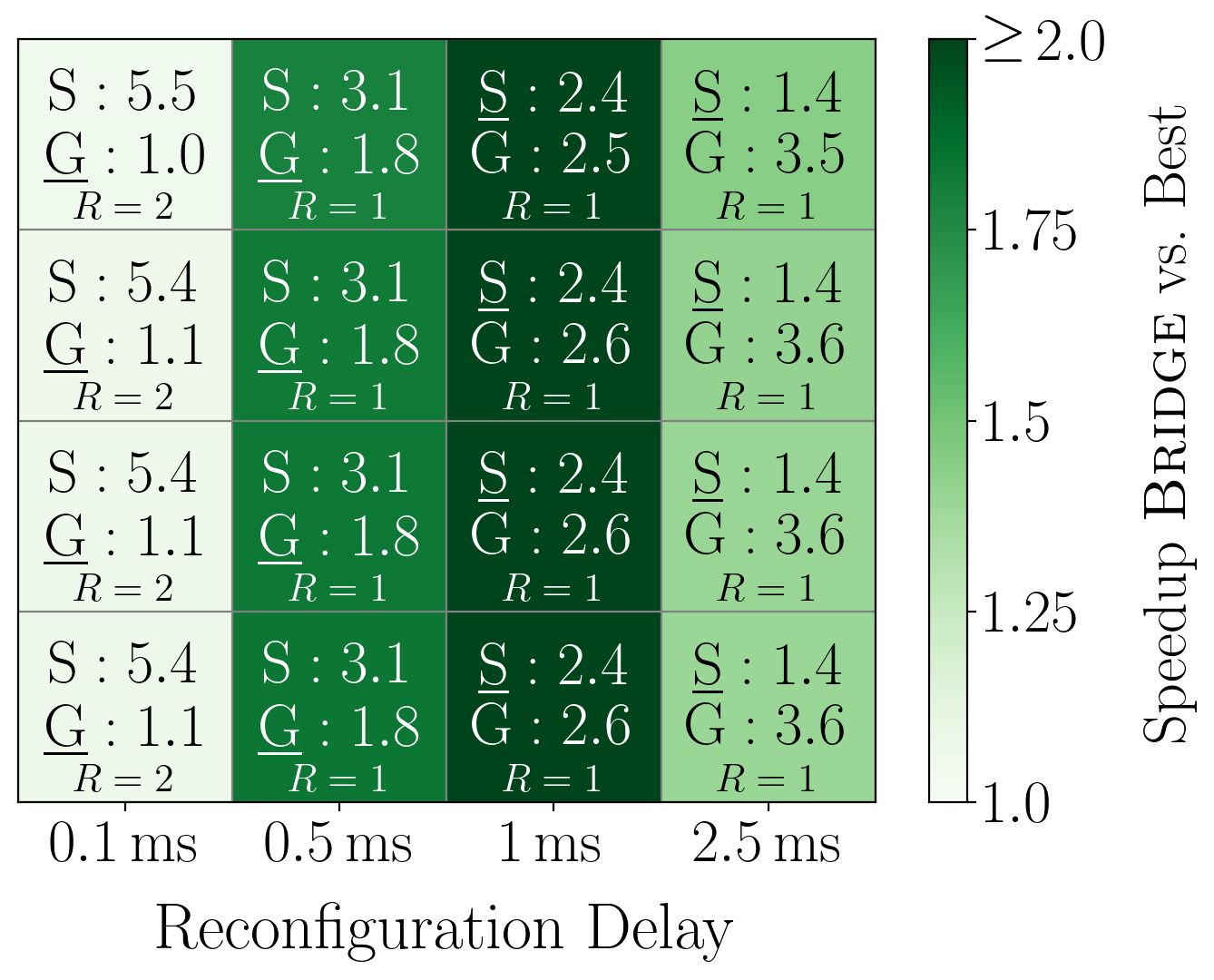}
        \caption{16 MB}
        \label{fig:matrix_all_hop_16mb}
    \end{subfigure}
    \caption{Speedup of \name compared to \sbruck and \gbruck for $n=64$ with varying per-hop delay.}
    \label{fig:matrix_all_hops}
        \vspace{-14px}
\end{figure}

In Figure~\ref{fig:matrix_all_hops}, we analyze \name under varying \textbf{per-hop delay} for representative small and large message sizes. Relative to \sbruck, \name becomes increasingly beneficial as per-hop communication grows, since reconfiguration can avoid longer multi-hop paths and better amortize the reconfiguration overhead against the higher communication costs. Compared to \gbruck, the opposite holds. Relative to both baselines, \name achieves the largest gains again in settings, where only sparse reconfigurations ($R=1$ or 2) are beneficial with speedups up to $2.3\times$. For \(m = 16\,\mathrm{MB}\) (Figure~\ref{fig:matrix_all_hop_16mb}), per-hop delay has no effect on the results.

\begin{figure}[t]
    \centering
    \begin{subfigure}[t]{0.495\columnwidth}
        \captionsetup{skip=2pt}
        \centering
        \includegraphics[width=\linewidth]{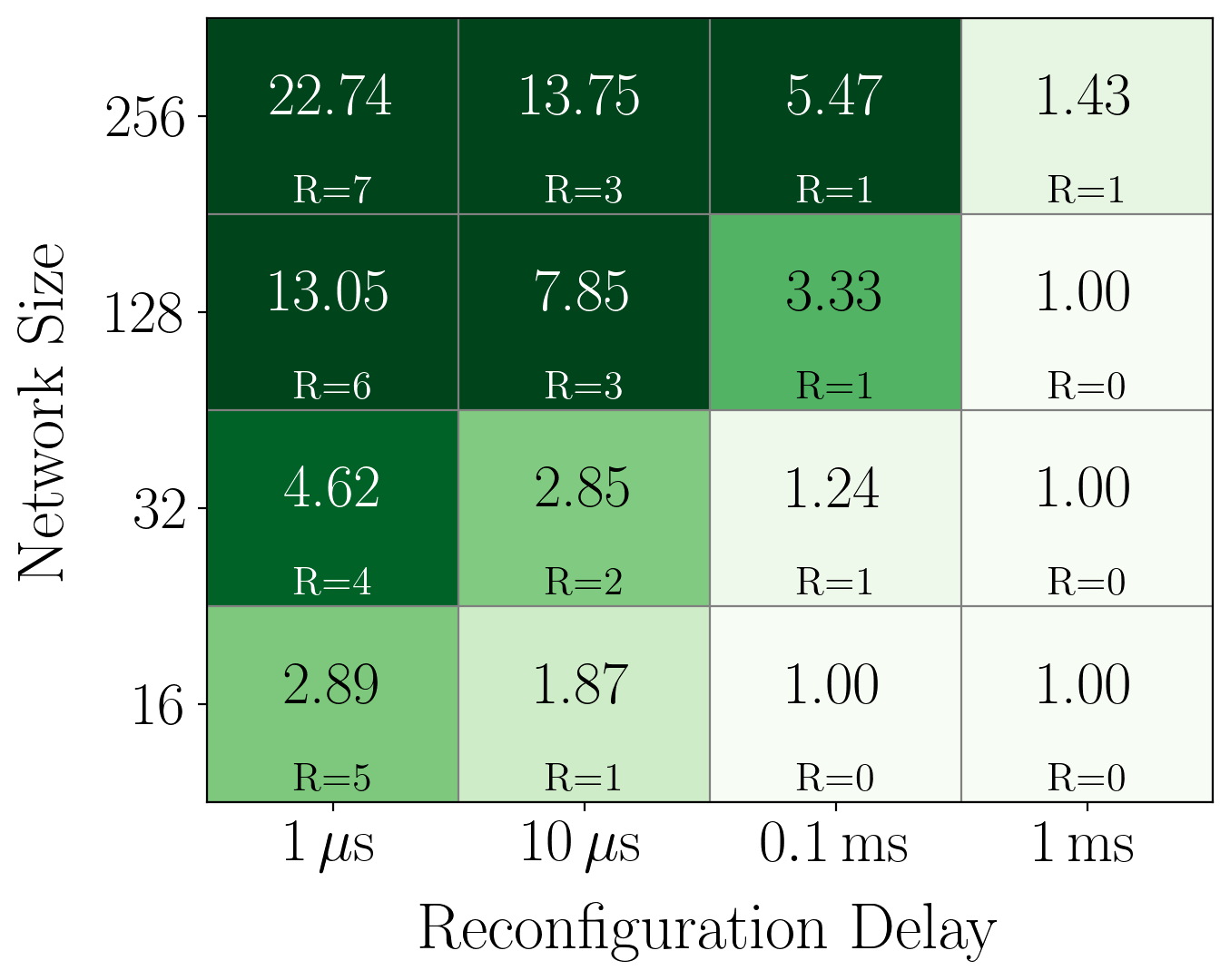}
        \caption{1 MB}
        \label{fig:matrix_all_nodes_1mb}
    \end{subfigure}
    \hfill
    \begin{subfigure}[t]{0.495\columnwidth}
        \captionsetup{skip=2pt}
        \centering
        \includegraphics[width=\linewidth]{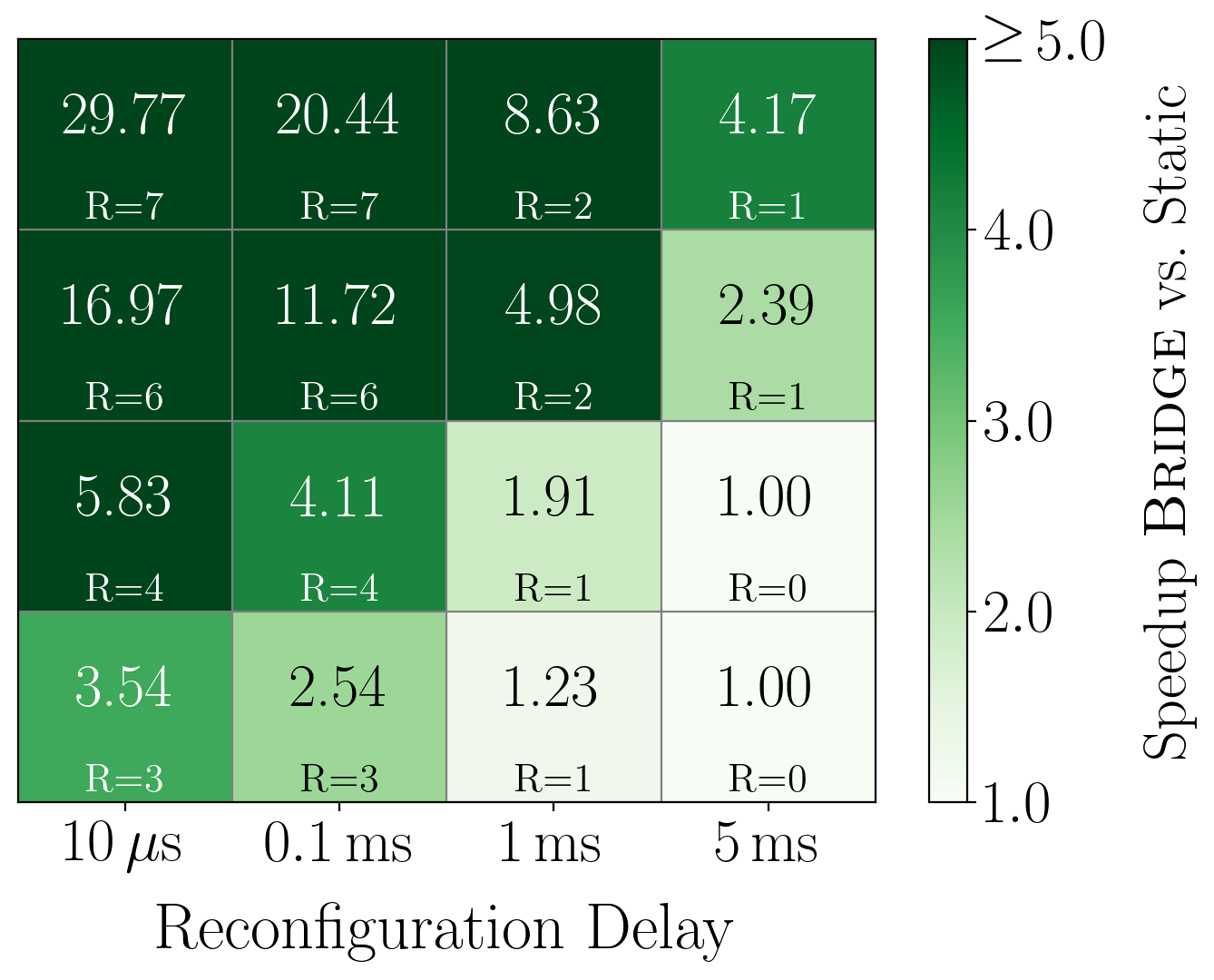}
        \caption{32 MB}
        \label{fig:matrix_all_nodes_32mb}
    \end{subfigure}
    \caption{Speedup of \name compared to \sbruck for 16 to 256 nodes with per hop delay $\alpha_h= 1\,\mu\mathrm{s}$.}
    \label{fig:matrix_all_nodes}
    \vspace{-14px}
\end{figure}

\begin{figure}[t]
    \centering
    \includegraphics[width=\columnwidth]{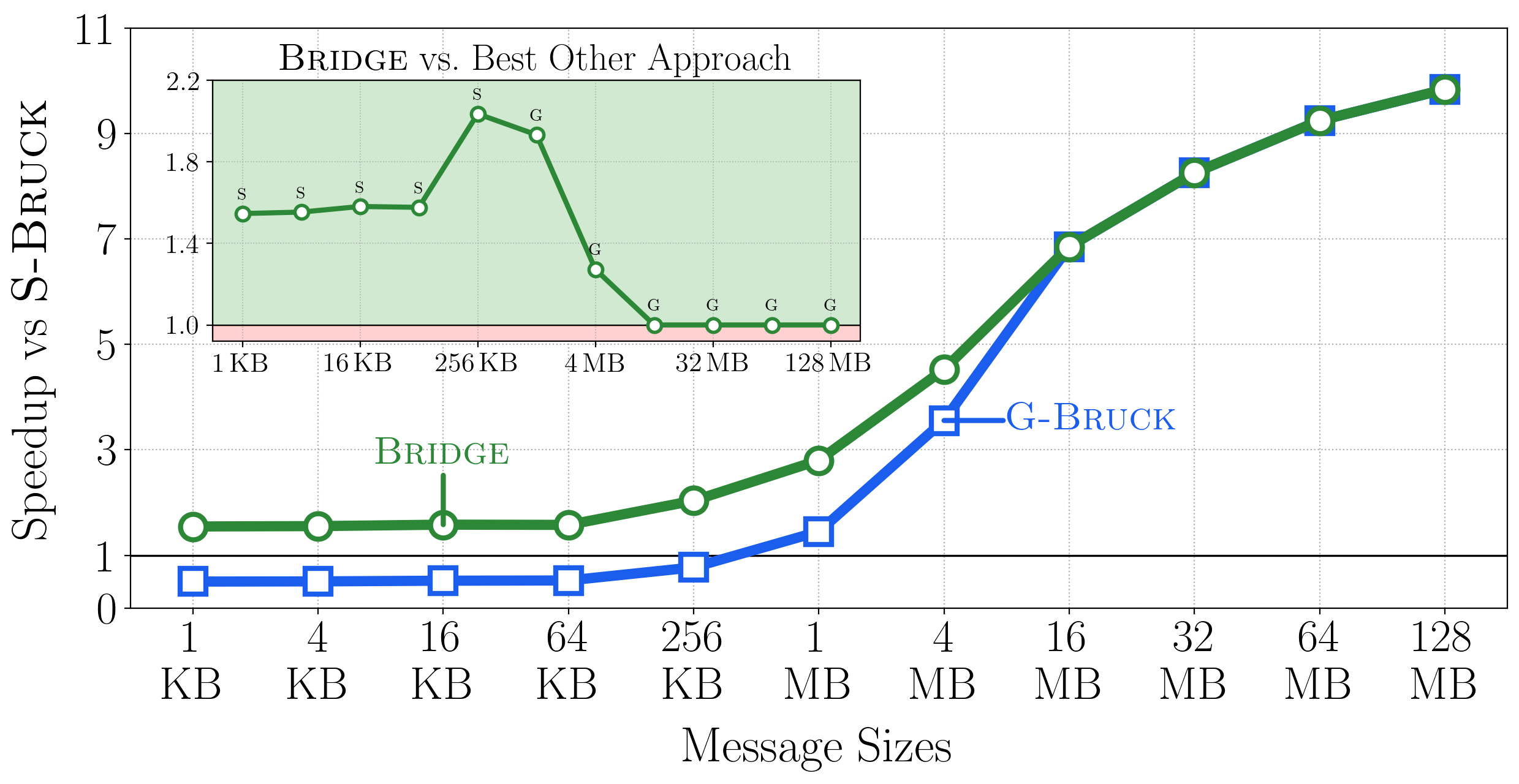}
    \caption{Speedup of \name and \gbruck against \sbruck for $n=64$ with reconfiguration delay of $10\,\mu\mathrm{s}$. The inset plot describes the improvement of \name over both baselines.}
  \label{fig:full_all_64}
  \vspace{-11px}
\end{figure}

To assess \name's performance across different \textbf{network sizes}, Figure~\ref{fig:matrix_all_nodes} reports the speedup for networks ranging from 16 to 256 nodes. We observe that \name achieves the largest gains in larger networks, particularly for large message sizes. For large networks, \name remains beneficial even at higher reconfiguration delays, achieving at least \(1.4\times\) speedup for 256 nodes across both message sizes. We next evaluate \name across the \textbf{full message-size range}. In Figure~\ref{fig:full_all_64}, we compare the speedup of \name and \gbruck over \sbruck for \(n=64\) and a reconfiguration delay of \(10\,\mu\)s, corresponding to the RotorNet configuration with 128 optical ports. The main plot shows that \name improves over \sbruck by \(1.4\times\) to \(3\times\) for small message sizes, with the speedup increasing to as much as \(10\times\) for large message sizes. For message sizes above 16\,MB, however, \gbruck performs identically, since reconfiguration at every step becomes beneficial in this regime. The inset plot highlights the advantage of \name over both baselines, showing that for this configuration \name achieves up to \(2.1\times\) speedup, first over \sbruck and then over \gbruck, for message sizes up to 4\,$\mathrm{MB}$.

\label{sec:ev:alltoall}

\textbf{How do message size, per-hop delay, and network size influence \name's performance gains?} 
We observe that \name achieves the largest gains over \sbruck for large message sizes, high per-hop delay, and large networks, as these increase communication cost and make topology reconfiguration more attractive. Compared to \gbruck, \name performs best in configurations opposite to those that favor gains over \sbruck, namely when the overhead of frequent reconfigurations outweighs its benefit.

\textbf{Under what conditions are reconfigurations feasible?
}
In order to answer this question, we look at the comparison to \sbruck. For message sizes above \(1\,\mathrm{MB}\), \name achieves speedups consistently of up to \(10.4\times\) over \sbruck. The benefit of \name also becomes consistent in larger networks, particularly from 128 nodes onward. Even under very small per-hop delays, \name still achieves speedups of up to \(5.4\times\), falling back to \sbruck only for small messages and high reconfiguration delays. Since reconfiguration delay typically scales with network size, configurations combining small networks with large \(\delta\) are of limited practical relevance. This shows that \name is beneficial whenever communication time exceeds the reconfiguration overhead and the network can be reconfigured with reasonably low delay.
\textbf{How does \name outperform existing approaches?}
\name outperforms \sbruck and \gbruck in complementary regimes: when \name outperforms \sbruck, it typically matches \gbruck, and vice versa. When $\delta$ is high, \name matches \sbruck at small message sizes and outperforms both baselines at larger sizes. At lower $\delta$, it outperforms both baselines at small message sizes and performs identically to \gbruck at larger sizes. As a result, the largest improvements over both baselines arise in configurations, where only sparse reconfigurations are worthwhile and \name can exploit \textit{reusable} links to reduce communication and \textit{amortize} the reconfiguration overhead across multiple steps.

\subsection{Reduce-Scatter}
\label{sec:ev:reduce}

In order to assess \name's performance for AllReduce, we consider several baselines, including a reconfigurable \halving strategy (\rehd), the Hamiltonian \ring algorithm, \sbruck, static \hd and \gbruck. In principle, the subrings induced by \bruck could also be used to enable OCS-link reuse under \hd, resulting in performance similar to \bruck (only with the reconfiguration strategy of \name). In the following, however, we consider \rehd as defined in prior work, i.e., starting from a ring topology and then applying BvN-based reconfigurations. Since \ring and \rehd combined outperform \sbruck and \gbruck for all workloads, we focus our comparison on the first two approaches. \ring serves as a strong baseline, as it avoids congestion and minimizes total data transmission, making it optimal for large message sizes.

\begin{figure}[t]
    \centering
    \begin{subfigure}[t]{0.495\columnwidth}
        \captionsetup{skip=2pt}
        \centering
        \includegraphics[width=\linewidth]{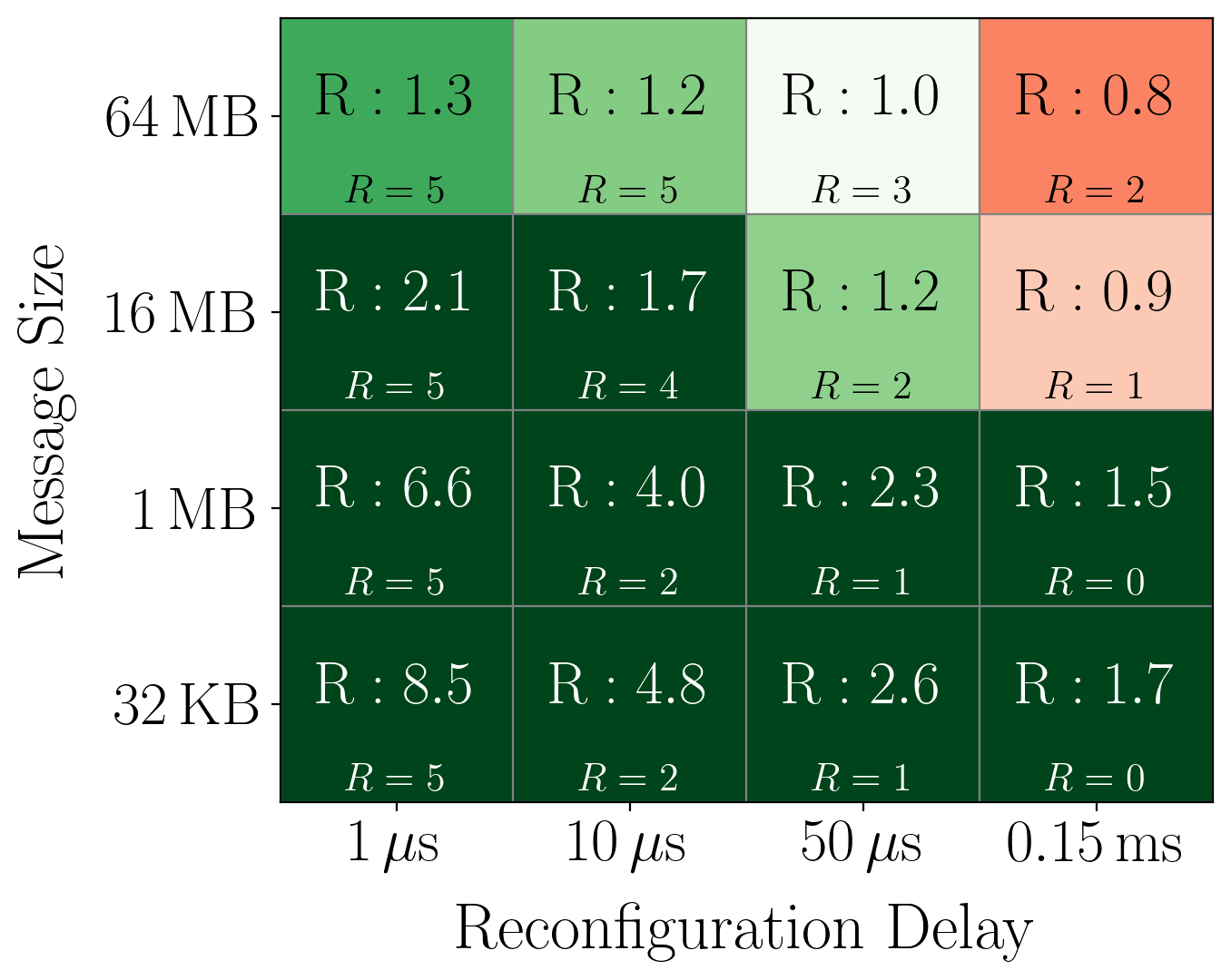}
        \caption{vs. \ring}
        \label{fig:matrix_reduce_msg_ring}
    \end{subfigure}
    \hfill
    \begin{subfigure}[t]{0.495\columnwidth}
        \captionsetup{skip=2pt}
        \centering
        \includegraphics[width=\linewidth]{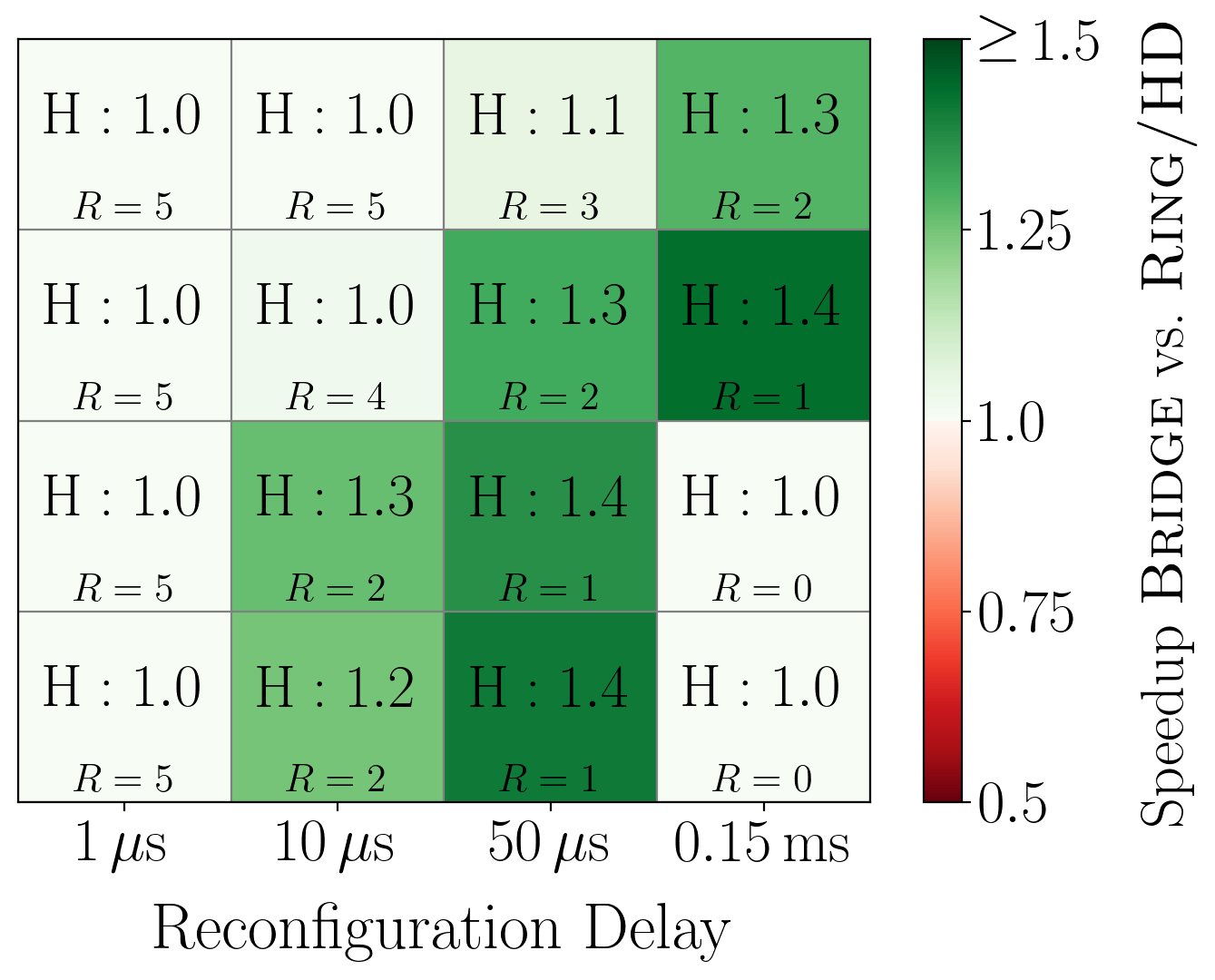}
        \caption{vs. \rehd}
        \label{fig:matrix_reduce_msg_hd}
    \end{subfigure}
    \caption{Speedup of \name compared to \ring/\rehd for $\alpha_h= 1\mu\mathrm{s}$, $b = 800\,\mathrm{Gbps}$ and $n=64$ with varying message size. }
    \label{fig:matrix_reduce_msg}
        \vspace{-13px}
\end{figure}

\begin{figure}[t]
    \centering
    \begin{subfigure}[t]{0.495\columnwidth}
        \captionsetup{skip=2pt}
        \centering
        \includegraphics[width=\linewidth]{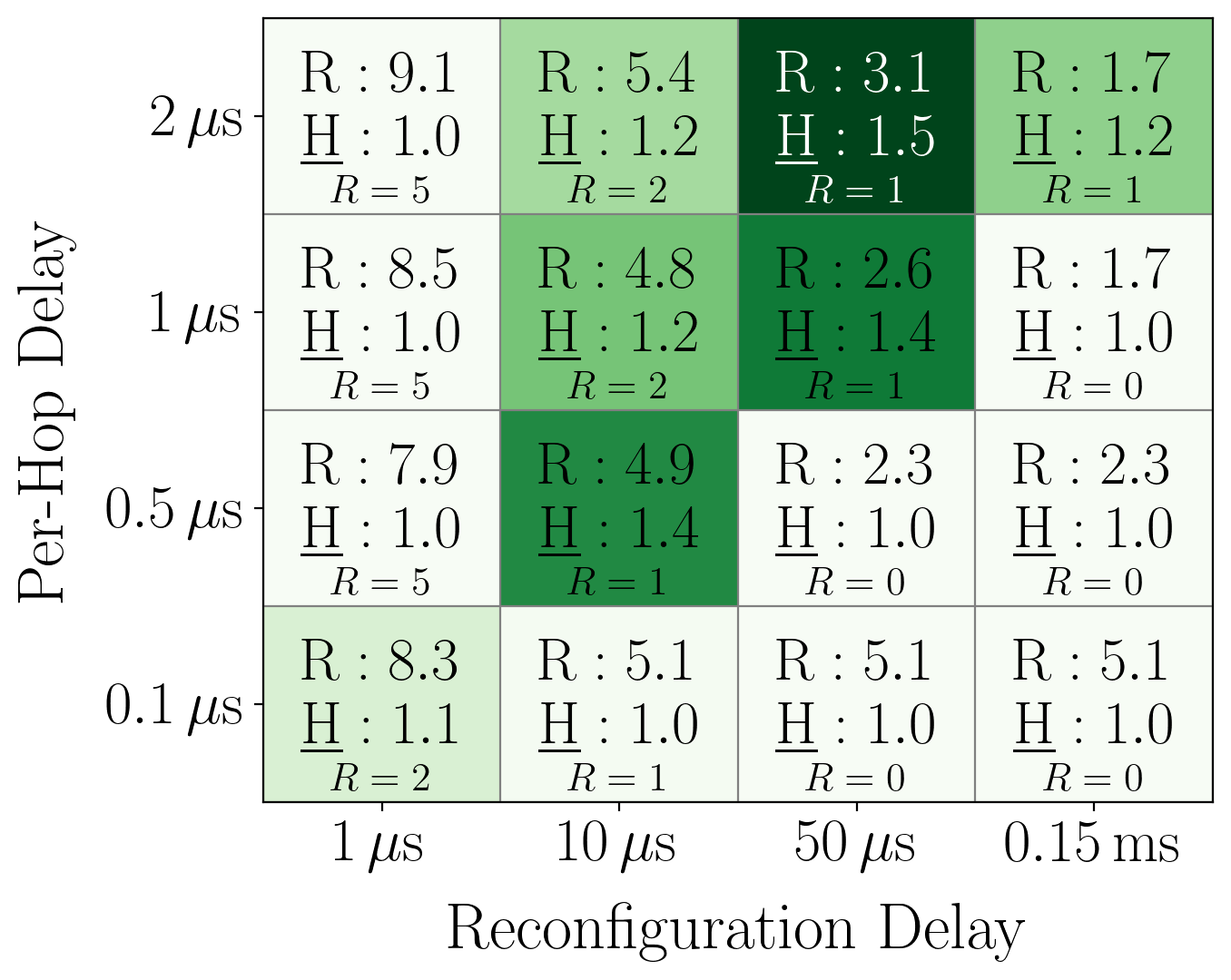}
        \caption{32 KB}
        \label{fig:matrix_reduce_hop_32kb}
    \end{subfigure}
    \hfill
    \begin{subfigure}[t]{0.495\columnwidth}
        \captionsetup{skip=2pt}
        \centering
        \includegraphics[width=\linewidth]{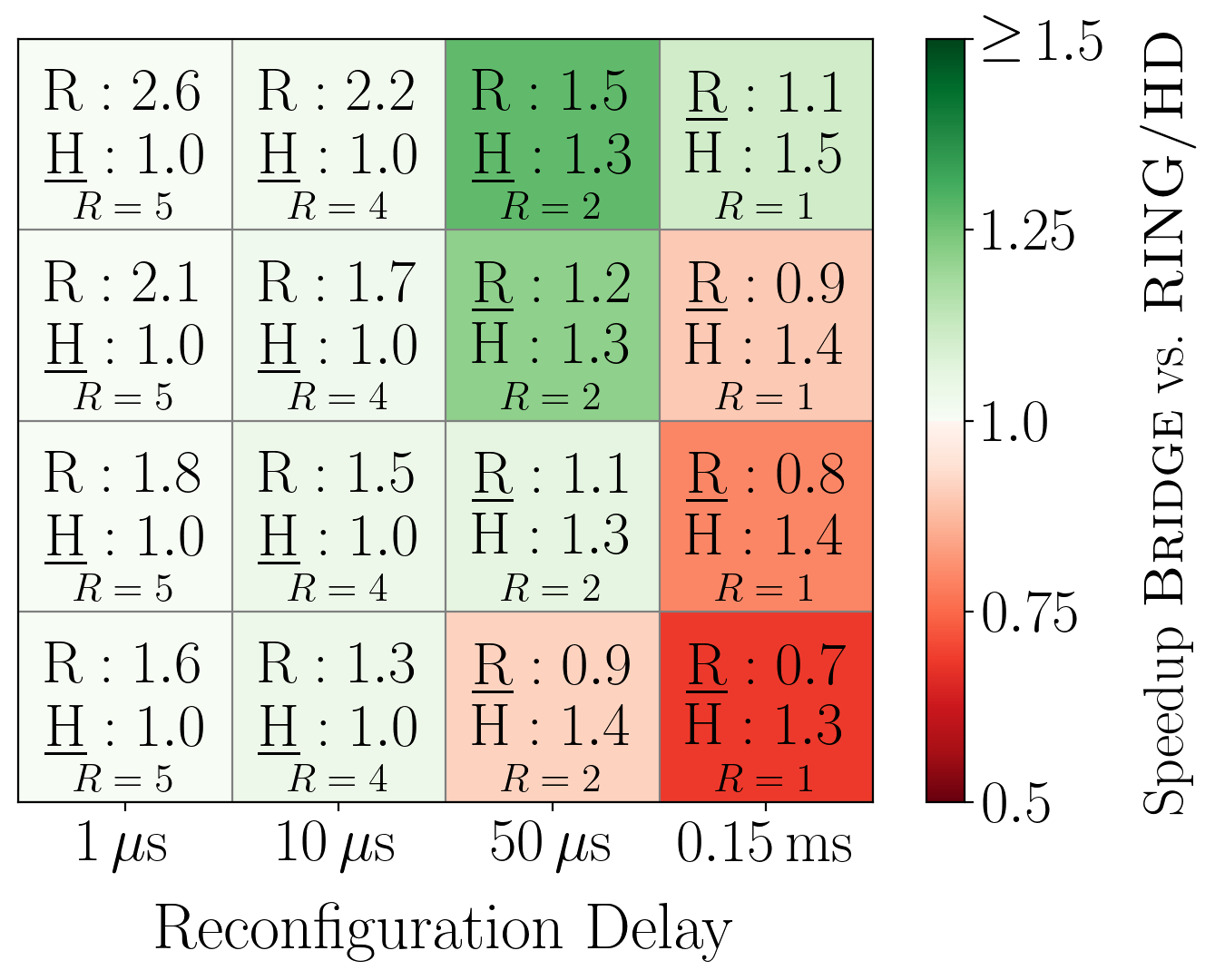}
        \caption{16 MB}
        \label{fig:matrix_reduce_hop_16mb}
    \end{subfigure}
    \caption{Speedup of \name compared to \ring and \rehd for $n=64$ with varying per-hop and reconfiguration delay.}
    \label{fig:matrix_reduce_hops}
        \vspace{-13px}
\end{figure}

\begin{figure}[t]
    \centering
    \begin{subfigure}[t]{0.495\columnwidth}
        \captionsetup{skip=2pt}
        \centering
        \includegraphics[width=\linewidth]{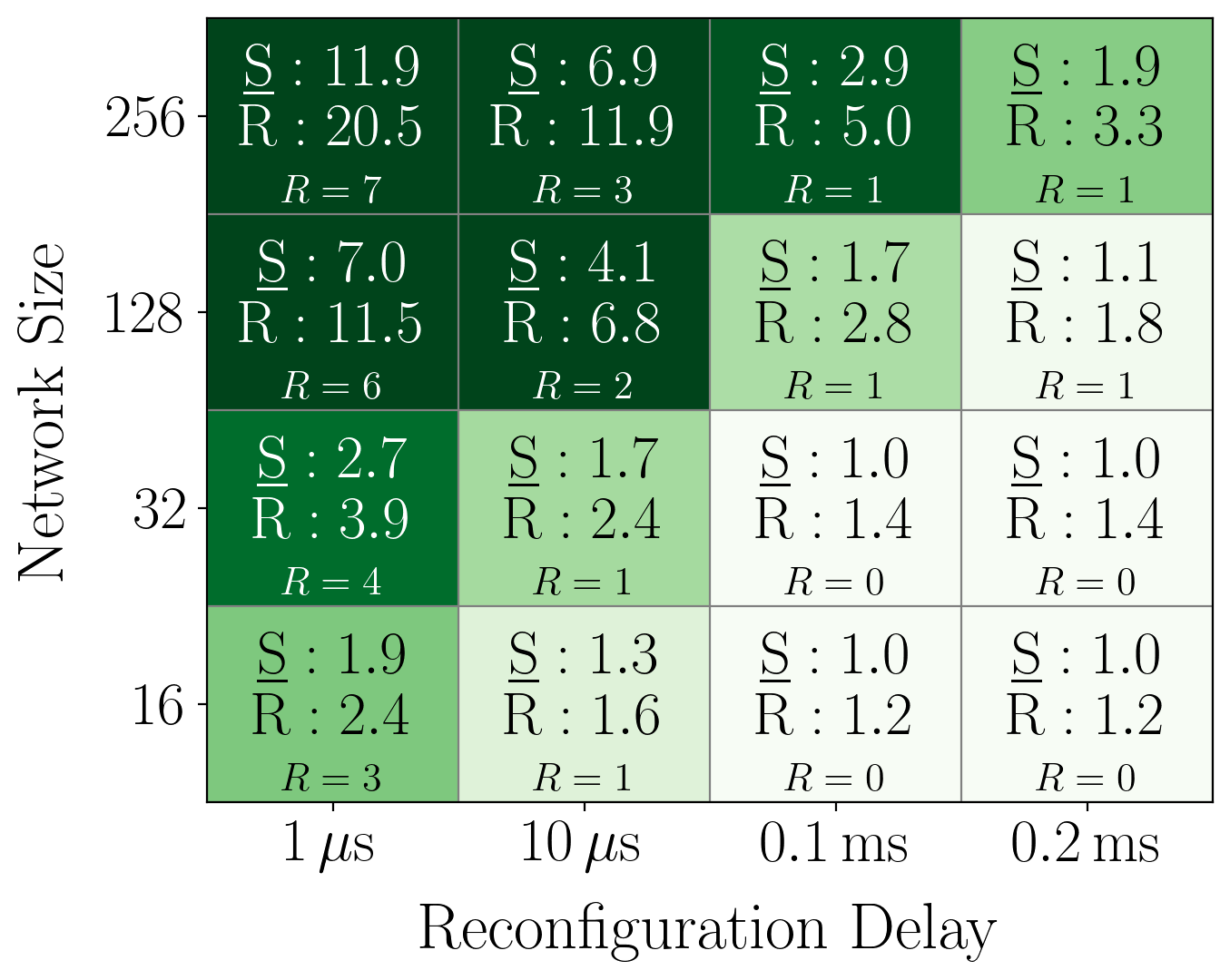}
        \caption{1 MB}
        \label{fig:matrix_reduce_nodes_1mb}
    \end{subfigure}
    \hfill
    \begin{subfigure}[t]{0.495\columnwidth}
        \captionsetup{skip=2pt}
        \centering
        \includegraphics[width=\linewidth]{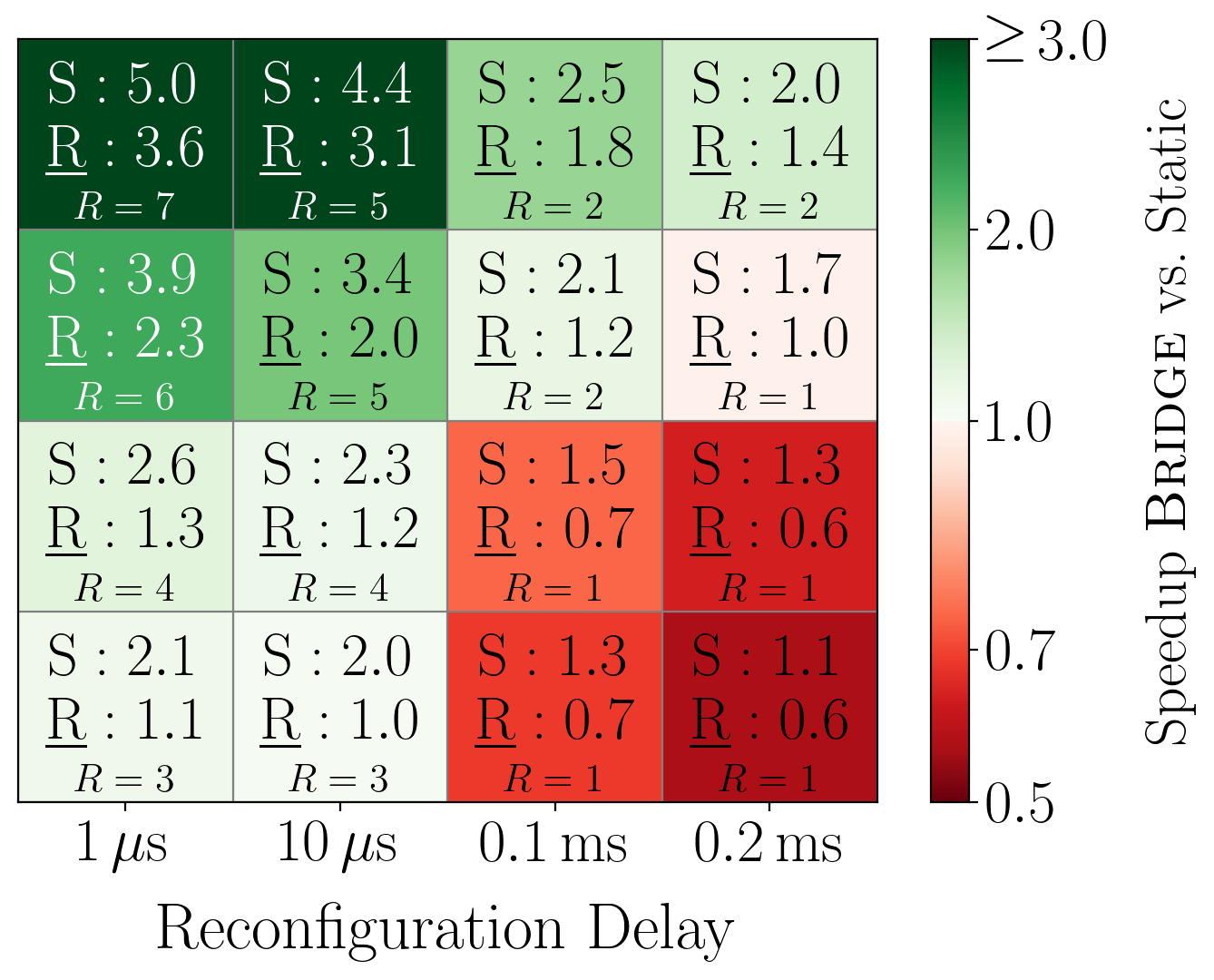}
        \caption{32 MB}
        \label{fig:matrix_reduce_nodes_32mb}
    \end{subfigure}
    \caption{Speedup of \name compared to \sbruck/\ring for networks of size 16 to 256 nodes with $\alpha_h= 1\mu\mathrm{s}$, $b = 800\,\mathrm{Gbps}$.}
    \label{fig:matrix_reduce_nodes}
    \vspace{-13px}
\end{figure}

\begin{figure}[t]
    \centering
    \includegraphics[width=\columnwidth]{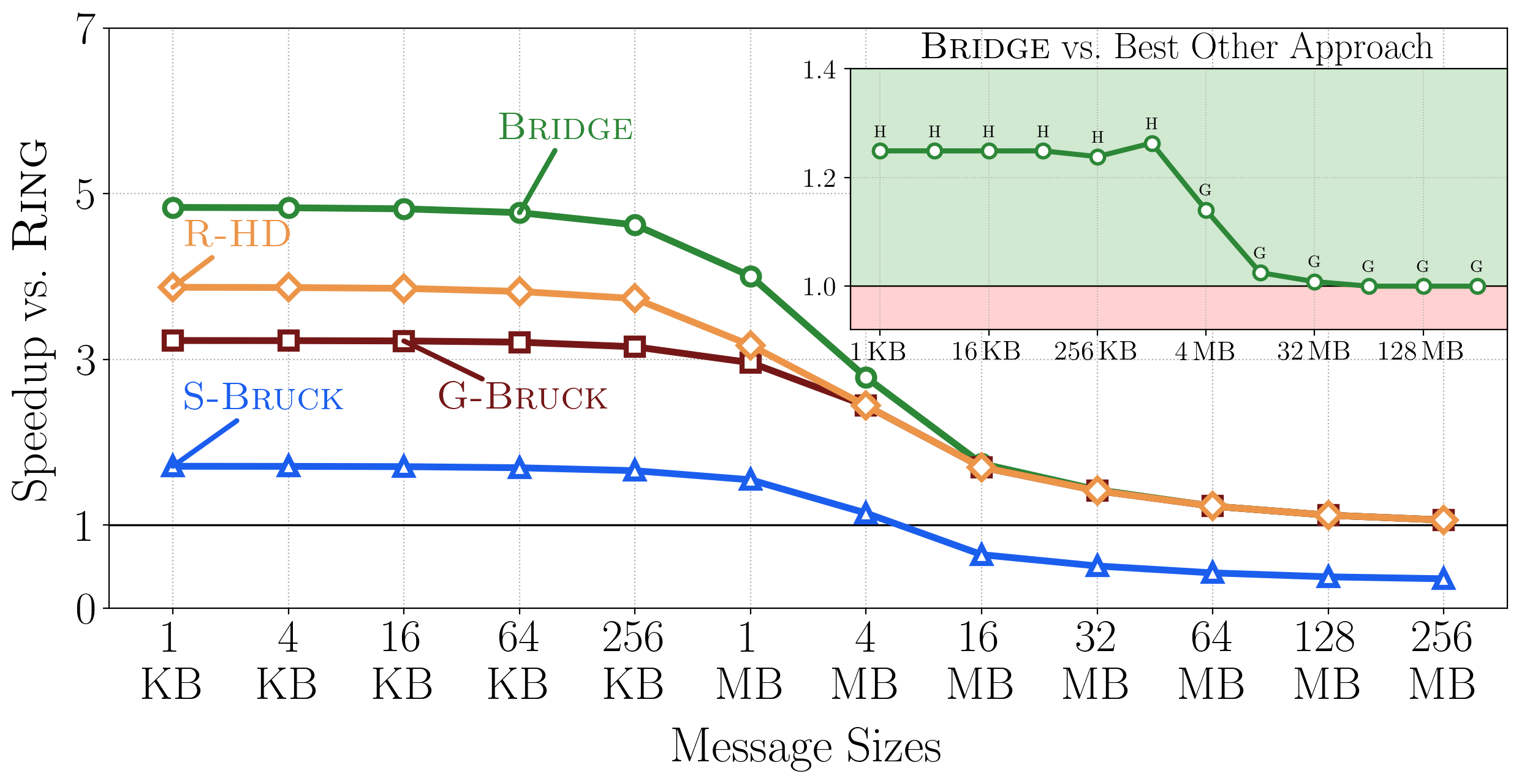}
  \caption{Speedup of all approaches compared to \ring for $n=64$ with $\delta = 10\,\mu\mathrm{s}$, $\alpha_h = 1\,\mu\mathrm{s}, b=800\,\mathrm{Gbps}$. The inset plot shows \name's improvement over the best baseline.}
  \label{fig:full_reduce_64}
    \vspace{-20px}
\end{figure}

Figure~\ref{fig:matrix_reduce_msg_ring} analyzes the speedup in completion time for varying \textbf{message sizes} against \ring and \rehd. Relative to \ring, \name achieves consistent speedups of up to \(8.5\times\) for small message sizes, whereas for large message sizes, this advantage diminishes: the speedup drops to around \(1.3\times\) at small reconfiguration delays, and for \(\delta = 0.15\,\mathrm{ms}\) \ring begins to outperform \name. At moderate message sizes, \name consistently outperforms \rehd by up to \(1.4\times\) in settings where one or two reconfigurations are beneficial. With varying \textbf{per-hop delay}, Figure~\ref{fig:matrix_reduce_hops} shows trends similar to All-to-All: higher per-hop delay makes reconfiguration more attractive. However, unlike All-to-All, the benefit is more limited because \ring remains highly competitive for large message sizes. Figure~\ref{fig:matrix_reduce_nodes} evaluates \textbf{network sizes} from 16 to 256 nodes to analyze the feasibility of \name against static baselines (stronger baseline is underlined). For small message sizes, \name becomes beneficial either in larger networks or at small reconfiguration delays, improving over static baselines by up to \(11.9\times\). For \(32\,\mathrm{MB}\), the benefit is concentrated in larger networks with low reconfiguration delay, where \name achieves speedups of \(3.6\times\). Figure~\ref{fig:full_reduce_64} shows the consistent trend that \name achieves its largest gains for small to medium message sizes outperforming \ring for messages up to 64\,$\mathrm{MB}$, providing speedups of up to \(5.0\times\). Further, \name outperforms all other baselines consistently; for large message sizes similar to \ring. In the same setting, \name improves over all existing approaches by up to \(1.3\times\) for message sizes up to 16\,$\mathrm{MB}$.

\textbf{How do message size, per-hop delay, and network size influence \name's performance gains?} 
The gains of \name are strongest when communication cost is dominated by per-hop delay (up to medium message sizes and large per-hop delays/networks). As message size grows, \ring becomes increasingly competitive as it already minimizes congestion. Likewise, higher per-hop delay and larger networks make reconfiguration more attractive, increasing the benefit of shortening multi-hop paths.

\textbf{Under what conditions are reconfigurations feasible?} 
For Reduce-Scatter, reconfigurations become beneficial when they reduce per-hop latency without incurring too much reconfiguration overhead. This is the case primarily for small to medium message sizes and in networks where multi-hop communication is sufficiently expensive (high $\alpha_h$ and $n$). By contrast, for large message sizes or high reconfiguration delays, \ring often remains the better choice, especially in smaller networks.

\textbf{How does \name outperform existing approaches?} 
\name outperforms \ring most clearly for small to medium message sizes, where shortcutting communication paths substantially reduces latency, yielding speedups of up to \(6.6\times\). Relative to \rehd, the gains are smaller but \name performs consistently better while reaching \(1.5\times\) speedups in settings where only one or two reconfigurations are beneficial. This shows that similar to All-to-All, \name benefits over existing approaches where sparse reconfigurations are most effective because they \textit{reuse} optical links to minimize latency cost and \textit{amortize} reconfiguration overhead across multiple steps.

\noindent \textbf{Summary:} \name is beneficial precisely in regimes where Figure~\ref{fig:costdistr} identifies optimization potential: when multi-hop latency and congestion contribute substantially to completion time, sparse reconfigurations can shorten communication paths and amortize their overhead across multiple subsequent steps. This is particularly effective for All-to-All and for latency-sensitive AllReduce workloads. In contrast, when transmission delay dominates in AllReduce, the room for improvement through topology adaptation is inherently limited. Thus, \name directly targets the optimization opportunity identified earlier by converting reusable optical shortcuts into workload-dependent completion-time gains.

\section{Discussion and Future Work}
\label{sec:discussion}
Our results show that \name is the most effective approach in network settings where sparse reconfigurations are beneficial. To extend \name to \textbf{reconfigurable multidimensional networks} such as tori~\cite{Jouppi2023TPUv4}, each node would have an OCS degree greater than two. Thereby, the network forms multiple subrings, which together form small \textit{subtori}. However, high-port-count OCS technologies have significantly higher reconfiguration delays, therefore this approach is likely practical only for small networks~\cite{Khani2021SiP}. Since \bruck's communication pattern is cyclic, it naturally enables \textbf{multiport capabilities} in reconfigurable networks, as each link is used in only one direction. Since OCS connections are bidirectional, a mirrored collective can be performed in parallel, as proposed for static collective algorithms in prior work~\cite{Daniele2024Swing, juerss2026}. For the architecture considered in this work, this yields a \(2\times\) transmission delay speedup and applies equally to \textsc{Ring}, \textsc{HD}, \textsc{S-Bruck}, and \textsc{G-Bruck}. The techniques and conditions proposed in this paper can also be applied to other \textbf{collective communication algorithms} with transitive and cyclic patterns, provided that their structure is compatible with the underlying ORN. One example is Trivance, which uses bidirectional communication to complete AllReduce in $\log_3 n$ steps~\cite{juerss2026}. Since its communication pattern is transitive, reconfigurations naturally form reusable subrings whose sizes decrease in powers of three~\cite{Juerss2026RevisitingBruck}. This generalization, however, is restricted to network sizes that are powers of three. For \textbf{All-to-Allv}, the reusable subring structure of \name remains applicable, but skewed traffic volumes may change which reconfigurations are worthwhile. Extending \name to traffic-aware reusable schedules that explicitly account for heterogeneous communication volumes remains future work.

\section{Conclusion}
We explore the design space of network reconfiguration during collective operations. We show that \textit{reusing} reconfigurable subrings across steps \textit{amortizes} reconfiguration overhead, reducing completion time and enabling efficient sparse reconfigurations even with millisecond-scale costs. We present a framework for identifying optimal schedules for \bruck's algorithm, and formalize this tradeoff for AllReduce and All-to-All. Overall, \name delivers substantial performance improvements for All-to-All completion time by typically \(3\times\) to \(10\times\) over static baselines and remains effective even with millisecond-scale reconfiguration delays. \name achieves speedups for AllReduce of up to \(1.5\times\) against existing reconfiguration strategies, and by $1.5\times$ to $6.6\times$ against the \ring algorithm on low to moderate-sized workloads. An important direction is to extend \name from offline schedule synthesis to an online runtime that adapts sparse reconfiguration to dynamic workloads. It remains open to validate these gains in a hardware implementation.

\medskip

\noindent \textbf{{\large Acknowledgments}}

\medskip
\noindent We thank Volker Stocker for valuable discussions and his feedback.  This work is part of a project that has received funding from the German Federal Ministry  of  Research,  Technology  and  Space  (BMFTR)  under grant 16DII141 “Weizenbaum Institut für die vernetzte Gesellschaft” and from the German Research Foundation (DFG)---project ReNO-2 of SPP 2378 (grant 511099228), 2025--2029.

\bibliographystyle{ACM-Reference-Format}
\bibliography{sample-base}

@inproceedings{Jouppi2023TPUv4,
author = {Jouppi, Norm and Kurian, George and Li, Sheng and Ma, Peter and Nagarajan, Rahul and Nai, Lifeng and Patil, Nishant and Subramanian, Suvinay and Swing, Andy and Towles, Brian and Young, Clifford and Zhou, Xiang and Zhou, Zongwei and Patterson, David A},
title = {TPU v4: An Optically Reconfigurable Supercomputer for Machine Learning with Hardware Support for Embeddings},
year = {2023},
isbn = {9798400700958},
publisher = {Association for Computing Machinery},
address = {New York, NY, USA},
url = {https://doi.org/10.1145/3579371.3589350},
doi = {10.1145/3579371.3589350},
booktitle = {Proceedings of the 50th Annual International Symposium on Computer Architecture},
articleno = {82},
numpages = {14},
location = {Orlando, FL, USA},
series = {ISCA '23}
}

@inproceedings{Ding2025Rails,
author = {Ding, Eric and Ouyang, Chuhan and Singh, Rachee},
title = {Photonic Rails in ML Datacenters},
year = {2025},
isbn = {9798400722806},
publisher = {Association for Computing Machinery},
address = {New York, NY, USA},
url = {https://doi.org/10.1145/3772356.3772414},
doi = {10.1145/3772356.3772414},
booktitle = {Proceedings of the 24th ACM Workshop on Hot Topics in Networks},
pages = {149–159},
numpages = {11},
location = {College Park, MD, USA},
series = {HotNets '25}
}

@inproceedings{Bruck94,
author = {Bruck, Jehoshua and Ho, Ching-Tien and Kipnis, Shlomo and Weathersby, Derrick},
title = {Efficient algorithms for all-to-all communications in multi-port message-passing systems},
year = {1994},
isbn = {0897916719},
publisher = {Association for Computing Machinery},
address = {New York, NY, USA},
url = {https://doi.org/10.1145/181014.181756},
doi = {10.1145/181014.181756},
booktitle = {Proceedings of the Sixth Annual ACM Symposium on Parallel Algorithms and Architectures},
pages = {298–309},
numpages = {12},
location = {Cape May, New Jersey, USA},
series = {SPAA '94}
}

@inproceedings{Qian2024Alibaba,
author = {Qian, Kun and Xi, Yongqing and Cao, Jiamin and Gao, Jiaqi and Xu, Yichi and Guan, Yu and Fu, Binzhang and Shi, Xuemei and Zhu, Fangbo and Miao, Rui and Wang, Chao and Wang, Peng and Zhang, Pengcheng and Zeng, Xianlong and Ruan, Eddie and Yao, Zhiping and Zhai, Ennan and Cai, Dennis},
title = {Alibaba HPN: A Data Center Network for Large Language Model Training},
year = {2024},
isbn = {9798400706141},
publisher = {Association for Computing Machinery},
address = {New York, NY, USA},
url = {https://doi.org/10.1145/3651890.3672265},
doi = {10.1145/3651890.3672265},
booktitle = {Proceedings of the ACM SIGCOMM 2024 Conference},
pages = {691–706},
numpages = {16},
location = {Sydney, NSW, Australia},
series = {ACM SIGCOMM '24}
}

@article{Thakur2005,
author = {Thakur, Rajeev and Rabenseifner, Rolf and Gropp, William},
title = {Optimization of Collective Communication Operations in MPICH},
year = {2005},
issue_date = {February  2005},
publisher = {Sage Publications, Inc.},
address = {USA},
volume = {19},
number = {1},
issn = {1094-3420},
url = {https://doi.org/10.1177/1094342005051521},
doi = {10.1177/1094342005051521},
journal = {Int. J. High Perform. Comput. Appl.},
month = feb,
pages = {49–66},
numpages = {18}
}

@INPROCEEDINGS{juerss2026,
author = {Juerss, Anton and Addanki, Vamsi and Schmid, Stefan},
title = {Trivance: Latency-Optimal AllReduce by Shortcutting Multiport Networks},
year = {2026},
isbn = {9798400724671},
publisher = {Association for Computing Machinery},
address = {New York, NY, USA},
url = {https://doi.org/10.1145/3789240.3829181},
booktitle = {Proceedings of the ACM SIGCOMM 2026 Conference},
pages = {160–176},
numpages = {17}
}

@inproceedings {Daniele2024Swing,
author = {Daniele De Sensi and Tommaso Bonato and David Saam and Torsten Hoefler},
title = {Swing: Short-cutting Rings for Higher Bandwidth Allreduce},
booktitle = {21st USENIX Symposium on Networked Systems Design and Implementation (NSDI 24)},
year = {2024},
isbn = {978-1-939133-39-7},
address = {Santa Clara, CA},
pages = {1445--1462},
url = {https://www.usenix.org/conference/nsdi24/presentation/de-sensi},
publisher = {USENIX Association},
month = apr
}

@article{hockney,
author = {Hockney, Roger W.},
title = {The communication challenge for MPP: Intel Paragon and Meiko CS-2},
year = {1994},
issue_date = {March 1994},
publisher = {Elsevier Science Publishers B. V.},
address = {NLD},
volume = {20},
number = {3},
issn = {0167-8191},
url = {https://doi.org/10.1016/S0167-8191(06)80021-9},
doi = {10.1016/S0167-8191(06)80021-9},
journal = {Parallel Comput.},
month = mar,
pages = {389–398},
numpages = {10}
}

@inproceedings{Qin2025,
author = {Qin, Le and Cui, Junwei and Cai, Weilin and Niu, Meng and Yang, Yan and Huang, Jiayi},
title = {Optimizing All-to-All Collective Communication with Fault Tolerance on Torus Networks},
year = {2025},
isbn = {9798400715730},
publisher = {Association for Computing Machinery},
address = {New York, NY, USA},
url = {https://doi.org/10.1145/3725843.3756057},
doi = {10.1145/3725843.3756057},
booktitle = {Proceedings of the 58th IEEE/ACM International Symposium on Microarchitecture},
pages = {659–674},
numpages = {16},
location = {Seoul, Korea},
series = {MICRO '25}
}

@inproceedings{Won2023Astrasim,
author = {Won, William and Heo, Taekyung and Rashidi, Saeed and Sridharan, Srinivas and Srinivasan, Sudarshan and Krishna, Tushar},
year = {2023},
month = {04},
pages = {283-294},
title = {ASTRA-sim2.0: Modeling Hierarchical Networks and Disaggregated Systems for Large-model Training at Scale},
doi = {10.1109/ISPASS57527.2023.00035}
}

@misc{ns3-simulator,
  title = {ns-3 Network Simulator}
}

@misc{NVIDIABlueField4Datasheet,
  author = {{NVIDIA}},
  title = {{NVIDIA BlueField-4 DPU Datasheet}}
}

@inproceedings {Aashaka2023TACCL,
author = {Aashaka Shah and Vijay Chidambaram and Meghan Cowan and Saeed Maleki and Madan Musuvathi and Todd Mytkowicz and Jacob Nelson and Olli Saarikivi and Rachee Singh},
title = {{TACCL}: Guiding Collective Algorithm Synthesis using Communication Sketches},
booktitle = {20th USENIX Symposium on Networked Systems Design and Implementation (NSDI 23)},
year = {2023},
isbn = {978-1-939133-33-5},
address = {Boston, MA},
pages = {593--612},
url = {https://www.usenix.org/conference/nsdi23/presentation/shah},
publisher = {USENIX Association},
month = apr
}

@inproceedings {Weiyang2023TopoOpt,
author = {Weiyang Wang and Moein Khazraee and Zhizhen Zhong and Manya Ghobadi and Zhihao Jia and Dheevatsa Mudigere and Ying Zhang and Anthony Kewitsch},
title = {{TopoOpt}: Co-optimizing Network Topology and Parallelization Strategy for Distributed Training Jobs},
booktitle = {20th USENIX Symposium on Networked Systems Design and Implementation (NSDI 23)},
year = {2023},
isbn = {978-1-939133-33-5},
address = {Boston, MA},
pages = {739--767},
url = {https://www.usenix.org/conference/nsdi23/presentation/wang-weiyang},
publisher = {USENIX Association},
month = apr
}

@inproceedings{Khani2021SiP,
author = {Khani, Mehrdad and Ghobadi, Manya and Alizadeh, Mohammad and Zhu, Ziyi and Glick, Madeleine and Bergman, Keren and Vahdat, Amin and Klenk, Benjamin and Ebrahimi, Eiman},
title = {SiP-ML: high-bandwidth optical network interconnects for machine learning training},
year = {2021},
isbn = {9781450383837},
publisher = {Association for Computing Machinery},
address = {New York, NY, USA},
url = {https://doi.org/10.1145/3452296.3472900},
doi = {10.1145/3452296.3472900},
booktitle = {Proceedings of the 2021 ACM SIGCOMM 2021 Conference},
pages = {657–675},
numpages = {19},
location = {Virtual Event, USA},
series = {SIGCOMM '21}
}

@inproceedings{Kumar2024Case,
author = {Kumar, Abhishek Vijaya and Devraj, Arjun and Bunandar, Darius and Singh, Rachee},
title = {A case for server-scale photonic connectivity},
year = {2024},
isbn = {9798400712722},
publisher = {Association for Computing Machinery},
address = {New York, NY, USA},
url = {https://doi.org/10.1145/3696348.3696856},
doi = {10.1145/3696348.3696856},
booktitle = {Proceedings of the 23rd ACM Workshop on Hot Topics in Networks},
pages = {290–299},
numpages = {10},
location = {Irvine, CA, USA},
series = {HotNets '24}
}

@inproceedings{Vamsi2025Bend,
author = {Addanki, Vamsi},
title = {When Light Bends to the Collective Will: A Theory and Vision for Adaptive Photonic Scale-up Domains},
year = {2025},
isbn = {9798400722806},
publisher = {Association for Computing Machinery},
address = {New York, NY, USA},
url = {https://doi.org/10.1145/3772356.3772395},
doi = {10.1145/3772356.3772395},
booktitle = {Proceedings of the 24th ACM Workshop on Hot Topics in Networks},
pages = {326–334},
numpages = {9},
location = {College Park, MD, USA},
series = {HotNets '25}
}

@article{Vamsi2023Mars,
author = {Addanki, Vamsi and Avin, Chen and Schmid, Stefan},
title = {Mars: Near-Optimal Throughput with Shallow Buffers in Reconfigurable Datacenter Networks},
year = {2023},
issue_date = {March 2023},
publisher = {Association for Computing Machinery},
address = {New York, NY, USA},
volume = {7},
number = {1},
url = {https://doi.org/10.1145/3579312},
doi = {10.1145/3579312},
journal = {Proc. ACM Meas. Anal. Comput. Syst.},
month = mar,
articleno = {2},
numpages = {43}
}

@misc{lightmatter2025passage,
  author = {{Lightmatter, Inc.}},
  title = {Passage Technology},
  year = {2025},
  url = {https://lightmatter.co/},
  urldate = {2026-04-21}
}

@misc{calient2022ocsdatasheet,
  author = {{CALIENT Technologies, Inc.}},
  title = {Calient's Optical Circuit Switch (S-Series) Datasheet},
  year = {2022},
  url = {https://www.calient.net/},
  urldate = {2026-04-21}
}

@inproceedings{Mellette2017RotorNet,
author = {Mellette, William M. and McGuinness, Rob and Roy, Arjun and Forencich, Alex and Papen, George and Snoeren, Alex C. and Porter, George},
title = {RotorNet: A Scalable, Low-complexity, Optical Datacenter Network},
year = {2017},
isbn = {9781450346535},
publisher = {Association for Computing Machinery},
address = {New York, NY, USA},
url = {https://doi.org/10.1145/3098822.3098838},
doi = {10.1145/3098822.3098838},
booktitle = {Proceedings of the Conference of the ACM Special Interest Group on Data Communication},
pages = {267–280},
numpages = {14},
location = {Los Angeles, CA, USA},
series = {SIGCOMM '17}
}

@inproceedings{Gangidi2024,
author = {Gangidi, Adithya and Miao, Rui and Zheng, Shengbao and Bondu, Sai Jayesh and Goes, Guilherme and Morsy, Hany and Puri, Rohit and Riftadi, Mohammad and Shetty, Ashmitha Jeevaraj and Yang, Jingyi and Zhang, Shuqiang and Fernandez, Mikel Jimenez and Gandham, Shashidhar and Zeng, Hongyi},
title = {RDMA over Ethernet for Distributed Training at Meta Scale},
year = {2024},
isbn = {9798400706141},
publisher = {Association for Computing Machinery},
address = {New York, NY, USA},
url = {https://doi.org/10.1145/3651890.3672233},
doi = {10.1145/3651890.3672233},
booktitle = {Proceedings of the ACM SIGCOMM 2024 Conference},
pages = {57–70},
numpages = {14},
location = {Sydney, NSW, Australia},
series = {ACM SIGCOMM '24}
}

@inproceedings{Amir2024,
author = {Amir, Daniel and Saran, Nitika and Wilson, Tegan and Kleinberg, Robert and Shrivastav, Vishal and Weatherspoon, Hakim},
title = {Shale: A Practical, Scalable Oblivious Reconfigurable Network},
year = {2024},
isbn = {9798400706141},
publisher = {Association for Computing Machinery},
address = {New York, NY, USA},
url = {https://doi.org/10.1145/3651890.3672248},
doi = {10.1145/3651890.3672248},
booktitle = {Proceedings of the ACM SIGCOMM 2024 Conference},
pages = {449–464},
numpages = {16},
location = {Sydney, NSW, Australia},
series = {ACM SIGCOMM '24}
}

@inproceedings{Zhu2015DCQCN,
author = {Zhu, Yibo and Eran, Haggai and Firestone, Daniel and Guo, Chuanxiong and Lipshteyn, Marina and Liron, Yehonatan and Padhye, Jitendra and Raindel, Shachar and Yahia, Mohamad Haj and Zhang, Ming},
title = {Congestion Control for Large-Scale RDMA Deployments},
year = {2015},
isbn = {9781450335423},
publisher = {Association for Computing Machinery},
address = {New York, NY, USA},
url = {https://doi.org/10.1145/2785956.2787484},
doi = {10.1145/2785956.2787484},
booktitle = {Proceedings of the 2015 ACM Conference on Special Interest Group on Data Communication},
pages = {523–536},
numpages = {14},
location = {London, United Kingdom},
series = {SIGCOMM '15}
}

@inproceedings{Porter2013,
author = {Porter, George and Strong, Richard and Farrington, Nathan and Forencich, Alex and Chen-Sun, Pang and Rosing, Tajana and Fainman, Yeshaiahu and Papen, George and Vahdat, Amin},
title = {Integrating microsecond circuit switching into the data center},
year = {2013},
isbn = {9781450320566},
publisher = {Association for Computing Machinery},
address = {New York, NY, USA},
url = {https://doi.org/10.1145/2486001.2486007},
doi = {10.1145/2486001.2486007},
booktitle = {Proceedings of the ACM SIGCOMM 2013 Conference on SIGCOMM},
pages = {447–458},
numpages = {12},
location = {Hong Kong, China},
series = {SIGCOMM '13}
}

@article{birkhoff1946three,
  author = {Birkhoff, Garrett},
  title = {Three observations on linear algebra},
  year = {1946},
  journal = {Univ. Nac. Tacuman, Rev. Ser. A}
}

@article{Avin2019DAN,
author = {Avin, Chen and Schmid, Stefan},
title = {Toward demand-aware networking: a theory for self-adjusting networks},
year = {2019},
issue_date = {October 2018},
publisher = {Association for Computing Machinery},
address = {New York, NY, USA},
volume = {48},
number = {5},
issn = {0146-4833},
url = {https://doi.org/10.1145/3310165.3310170},
doi = {10.1145/3310165.3310170},
journal = {SIGCOMM Comput. Commun. Rev.},
month = jan,
pages = {31–40},
numpages = {10}
}

@INPROCEEDINGS{hoefler2010toward,
author = {Hoefler, Torsten and Gropp, William and Thakur, Rajeev and Träff, Jesper},
year = {2010},
month = {09},
pages = {21-30},
title = {Toward Performance Models of MPI Implementations for Understanding Application Scaling Issues},
isbn = {978-3-642-15645-8},
doi = {10.1007/978-3-642-15646-5_3}
}

@INPROCEEDINGS{Hu2025NCCL,
  author={Hu, Zhiyi and Shen, Siyuan and Bonato, Tommaso and Jeaugey, Sylvain and Alexander, Cedell and Spada, Eric and Dinan, James and Hammond, Jeff and Hoefler, Torsten},
  booktitle={2025 IEEE Symposium on High-Performance Interconnects (HOTI)}, 
  title={Demystifying NCCL: An In-Depth Analysis of GPU Communication Protocols and Algorithms}, 
  year={2025},
  volume={},
  number={},
  pages={48-59},
  doi={10.1109/HOTI66940.2025.00024}}

@misc{reuters2026aiinfra,
  author = {{Reuters}},
  title = {Big Tech to invest about \$650 billion in AI in 2026, Bridgewater says},
  year = {2026},
  url = {https://www.reuters.com/business/big-tech-invest-about-650-billion-ai-2026-bridgewater-says-2026-02-23/},
  urldate = {2026-04-21}
}

@INPROCEEDINGS{Rahman2026Harvest,
author = {Rahman, Mahir and Joseph, Samuel and Kodkani, Nihar and Arzani, Behnaz and Addanki, Vamsi},
title = {Harvest: Adaptive Photonic Switching Schedules for Collective Communication in Scale-up Domains},
year = {2026},
isbn = {9798400724671},
publisher = {Association for Computing Machinery},
address = {New York, NY, USA},
url = {https://doi.org/10.1145/3789240.3829166},
booktitle = {Proceedings of the ACM SIGCOMM 2026 Conference},
pages = {1576–1600},
numpages = {25}
}

@misc{hammer2025,
  author = {Hammer, Sarah-Michelle and Juerss, Anton and Schmid, Stefan},
  title = {Shortcut: In-Collective Topology Reconfiguration for Low-Latency AllReduce},
  year = {2026},
  archivePrefix = {arXiv},
  primaryClass = {cs.NI},
  url = {https://arxiv.org/abs/2510.03491}
}

@inproceedings{Juerss2026RevisitingBruck,
author = {Juerss, Anton and Schmid, Stefan},
title = {Revisiting Bruck: Phase-Efficient All-to-All Collective Communication in Reconfigurable Networks},
year = {2026},
isbn = {9798400724671},
publisher = {Association for Computing Machinery},
address = {New York, NY, USA},
url = {https://doi.org/10.1145/3789240.3822611},
doi = {10.1145/3789240.3822611},
booktitle = {Proceedings of the ACM SIGCOMM 2026 Conference},
pages = {2142–2149},
numpages = {8},
location = {Denver, CO, USA},
series = {SIGCOMM '26}
}

@inproceedings{Juerss2027Sigmetrics,
  author = {Juerss, Anton and Krebs, Mitja Daniel and Melnyk, Darya and Schmid, Stefan},
  title = {Optimizing Hybrid Reconfigurable Datacenter Networks for All-to-All Collective Communications},
  year = {2027},
  booktitle = {SIGMETRICS '27}
}

@article{Zerwas2023Duo,
author = {Zerwas, Johannes and Gy{\"o}rgyi, Csaba and Blenk, Andreas and Schmid, Stefan and Avin, Chen},
title = {Duo: A High-Throughput Reconfigurable Datacenter Network Using Local Routing and Control},
year = {2023},
issue_date = {March 2023},
publisher = {Association for Computing Machinery},
address = {New York, NY, USA},
volume = {7},
number = {1},
url = {https://doi.org/10.1145/3579449},
doi = {10.1145/3579449},
journal = {Proc. ACM Meas. Anal. Comput. Syst.},
month = mar,
articleno = {20},
numpages = {25}
}

@article{Avin2025Rev,
author = {Avin, Chen and Schmid, Stefan},
title = {Revolutionizing Datacenter Networks via Reconfigurable Topologies},
year = {2025},
issue_date = {June 2025},
publisher = {Association for Computing Machinery},
address = {New York, NY, USA},
volume = {68},
number = {6},
issn = {0001-0782},
url = {https://doi.org/10.1145/3708980},
doi = {10.1145/3708980},
journal = {Commun. ACM},
month = jun,
pages = {44–53},
numpages = {10}
}

@article{traff2025,
author = {Tr{\"a}ff, Jesper Larsson},
title = {Optimal, Non-pipelined Reduce-scatter and Allreduce Algorithms with an Application to All-to-all Communication},
year = {2025},
issue_date = {December 2025},
publisher = {Association for Computing Machinery},
address = {New York, NY, USA},
volume = {12},
number = {4},
issn = {2329-4949},
url = {https://doi.org/10.1145/3760789},
doi = {10.1145/3760789},
journal = {ACM Trans. Parallel Comput.},
month = dec,
articleno = {10},
numpages = {23}
}

\end{document}